\documentclass[hidelinks, 12pt]{article}
\usepackage{geometry}
\usepackage[normalem]{ulem}
\usepackage{setspace}
\usepackage[utf8]{inputenc}
\usepackage{times}
\usepackage{epigraph} 
\usepackage{xcolor}
\usepackage{subcaption}
\usepackage{bm}
\usepackage{comment}
\usepackage{amsfonts}
\usepackage{amsmath}
\usepackage{amssymb}
\usepackage{amsthm}
\usepackage{graphicx, caption, subcaption}
\usepackage{xfrac}
\usepackage{multirow}
\usepackage{pgfplots}
\usepackage{tikz}
\usepackage{dcolumn}
\usepackage{threeparttable}
\usepackage{lscape}
\usepackage{tabularx}
\usepackage{booktabs}
\usepackage{caption}
\usepackage{siunitx}
\usepackage{adjustbox}
\usepackage{accents}
\usepackage{epstopdf}
\usepackage{hyperref}
\usepackage[noabbrev]{cleveref}
\usepackage{tasks}
\usepackage{appendix}
\usepackage{authblk}
\usepackage{mathtools}
\usepackage{fancyhdr}
\usepackage[round]{natbib}

\usepackage [english]{babel}

\usepackage{blindtext}

\usepackage{datetime}

\newdateformat{monthyeardate}{%
  \monthname[\THEMONTH], \THEYEAR}

\usepackage{float}
\usepackage{rotating}
\usepackage{enumitem,amssymb}
\usepackage{soul}
\usepackage{bbm}

\pgfplotsset{compat=1.15}

\usetikzlibrary{plotmarks}
\tikzset{cross/.style={cross out, draw=black, minimum size=2*(#1-\pgflinewidth), inner sep=0pt, outer sep=0pt},
cross/.default={1pt}}

\usepackage{footmisc}

\newtheorem{definition}{Definition}
\newtheorem{lemma}{Lemma}
\newtheorem{proposition}{Proposition}

\newtheorem{corollary}{Corollary}
\newtheorem{assumption}{Assumption}

\theoremstyle{definition}
\newtheorem{remark}{Remark}

\theoremstyle{definition}

\theoremstyle{definition}

\crefname{prop}{Proposition}{Propositions}
\crefname{mydef}{Definition}{Definitions}
\crefname{lemma}{Lemma}{Lemmas}
\crefname{definition}{Definition}{Definitions}
\crefname{theorem}{Theorem}{Theorems}
\crefname{proposition}{Proposition}{Propositions}
\crefname{corollary}{Corollary}{Corollaries}
\crefname{assumption}{Assumption}{Assumptions}
\crefname{remark}{Remark}{Remarks}
\crefname{claim}{Claim}{Claims}
\crefname{section}{Section}{Sections}
\crefname{figure}{Figure}{Figures}
\crefname{exmp}{Example}{Examples}
\crefname{observation}{Observation}{Observations}

\DeclareMathOperator{\E}{\mathbb{E}}
\newcommand{\reals}{\mathbb{R}}

\DeclareMathOperator*{\inte}{\text{int}}

\makeatletter
\DeclareRobustCommand\citepos													
  {\begingroup\def\NAT@nmfmt##1{{\NAT@up##1's}}%
   \NAT@swafalse\let\NAT@ctype\z@\NAT@partrue
   \@ifstar{\NAT@fulltrue\NAT@citetp}{\NAT@fullfalse\NAT@citetp}}

\pretocmd{\NAT@citex}{%
  \let\NAT@hyper@\NAT@hyper@citex
  \def\NAT@postnote{#2}%
  \setcounter{NAT@total@cites}{0}%
  \setcounter{NAT@count@cites}{0}%
  \forcsvlist{\stepcounter{NAT@total@cites}\@gobble}{#3}}{}{}
\newcounter{NAT@total@cites}
\newcounter{NAT@count@cites}
\def\NAT@postnote{}

\def\NAT@hyper@citex#1{
  \stepcounter{NAT@count@cites}%
  \hyper@natlinkstart{\@citeb\@extra@b@citeb}#1%
  \ifnumequal{\value{NAT@count@cites}}{\value{NAT@total@cites}}
    {\if*\NAT@postnote*\else\NAT@cmt\NAT@postnote\global\def\NAT@postnote{}\fi}{}%
  \ifNAT@swa\else\if\relax\NAT@date\relax
  \else\NAT@@close\global\let\NAT@nm\@empty\fi\fi								
  \hyper@natlinkend}
\renewcommand\hyper@natlinkbreak[2]{#1}

\patchcmd{\NAT@citex}															
  {\ifNAT@swa\else\if*#2*\else\NAT@cmt#2\fi
   \if\relax\NAT@date\relax\else\NAT@@close\fi\fi}{}{}{}
\patchcmd{\NAT@citex}
  {\if\relax\NAT@date\relax\NAT@def@citea\else\NAT@def@citea@close\fi}
  {\if\relax\NAT@date\relax\NAT@def@citea\else\NAT@def@citea@space\fi}{}{}
\patchcmd{\NAT@cite}{\if*#3*}{\if*\NAT@postnote*}{}{}
\makeatother

\title{Simple Communication}
\author{Jacopo Bizzotto \quad Nathan Hancart\thanks{Bizzotto: Oslo Metropolitan University, Hancart: University of Oslo. We are grateful to Bård Harstad, Bart Lipman, Luca Onnis and Ran Spiegler for useful feedback and suggestions. We also thank Katinka Holtsmark for early discussions on this project.}}
\date{\monthyeardate\today}

\begin{document}
\maketitle

\onehalfspacing
\begin{abstract}
\noindent 
We study simple communication in a multidimensional cheap talk model with aligned preferences.
An expert communicates with a decision-maker using a score, i.e., an aggregator of a multidimensional state into a one-dimensional message. Even though the expert and the decision-maker share the same payoffs, equilibrium payoffs are lower than those achievable under commitment to a score. Under quadratic-loss utility, any equilibrium  score  must be linear in the state or a coarsening of a linear score.    
Finally, for normally distributed states, we characterize the equilibrium linear scores and show they include the ex-ante best and worst linear scores.
\end{abstract}

\newpage

\section{Introduction}

In a December 2020 interview, the epidemiologist Caroline Buckee joked that she wanted to print T-shirts reading “\textsl{Wear a mask. Indoor dining is a bad idea. No large gatherings}”. Experts often face the tension captured in this remark: advice has to be simple even when the matters it addresses are complex.
Similarly, an economist advising a politician cannot fully convey all the heterogeneous effects of an economic shock on consumers across socioeconomic backgrounds and regions. Instead the expert must offer broad assessments and aggregate statistics.

In this paper we study simple communication on complex issues.
In the examples above, the complexity of the issues stems from their multidimensionality, and simplicity is achieved by 
 keeping messages low-dimensional. The epidemiologist gives one piece of broad advice, instead of giving multiple ones tailored to each segment of the population. The economist aggregates the heterogeneous impacts of the shock into few general assessments.

We model communication as a cheap talk game. A sender observes a multidimensional state of the world and sends a message to a receiver. The receiver takes a multidimensional action to minimize a quadratic loss function. Sender and receiver share the same payoffs. We  require that the sender's \textsl{equilibrium strategy}  maps the state  to a one-dimensional message, and satisfies a weak form of continuity we dub Intermediate Value Property (IVP). We refer to such a strategy as a \textsl{score}.

In line with our examples, we model complexity through the multidimensional state and we model simple communications through scores. For the epidemiologist,  the one-dimensional message might correspond to a single piece of advice on the degree of social distancing to the whole population. For the economist, the message could be a single assessment of the size of the shock or the relative impact on different populations.\footnote{Other examples include critics’ reviews, such as restaurant or movie ``star ratings,’’ which compress a rich experience into a single number, and short briefings written for time-pressed managers.}
Only requiring one-dimensional messages would not, however, be enough to capture simple communication. Since injections from $\reals^n$ to $\reals$ exist, the sender could fully reveal the state. We additionally require that the sender's strategy satisfies the IVP. This restriction imposes natural restrictions and prevents injections. Roughly speaking, the IVP requires that small changes in the state cause at most small changes in the score.
The property does not impose monotonicity or other functional-form restrictions and accommodates both continuous and discrete images.\footnote{Every continuous sender strategy satisfies the IVP. If a strategy satisfying the IVP takes values in a discrete set, such as ``small'', ``moderate'' and ``large'' categories for an economic shock, then a marginal change in the state moves the score by at most one category.}

We model a sender who is able to adopt any communication strategy, including complex ones, yet we  require communication to be simple in equilibrium. 
This is also in line with our examples. Public policy experts are able to lay out elaborate plans of action, with provisions for many possible contingencies and prescriptions tailored to different risk profiles. To reach the general public, however, effective recommendations must be simple and short.

We show  that, despite the players' aligned preferences, equilibrium payoffs are strictly lower than those achievable under commitment to a score (\cref{proposition:calendar}).\footnote{\cref{prop:existence} establishes that an equilibrium score always exists.} In other words, for any loss function, there exist scores that ensure higher payoffs than every equilibrium score. The intuition is as follows. We propose an equilibrium-selection criterion rather than a restriction on the strategy space; once the state is observed, the sender remains free to deviate to any message. A score that yields high payoffs ex-ante might not be an equilibrium score if the sender has a profitable deviation in some states. In \cref{sec:2X2}, we illustrate the tension between informativeness and equilibrium behavior in a simplified setting.

Having shown that equilibrium scores are suboptimal, we next characterize the restrictions that incentive compatibility places on their shape. In \cref{prop:linear}, we show that  every equilibrium score is payoff-equivalent to a linear score or a coarsening of one,\footnote{In other words, any equilibrium score must be quasilinear, in the sense of both quasiconvex and quasiconcave.} and  the posterior expectations are collinear. The latter observation implies, for instance, that focusing on two dimensions of the state, a higher message must induce a higher expectation in both dimensions or a higher expectation in one and lower in the other. 

Suppose the state is the impact of a macroeconomic shock on low and high income households. \cref{prop:linear} implies that the economist communicates by taking a weighted average of the two effects, and possibly coarsens that message further. Moreover, a higher message must indicate a stronger impact of the shock on both groups, or a stronger relative impact on one group versus the other. In particular, the score cannot convey a non-monotonic relation of the impact between the two groups.

With additional structure on the distribution of states, we can further characterize equilibrium scores. When the state is normally distributed, the best equilibrium score is linear and coincides with the ex-ante optimal linear score (\cref{prop:normal}). There are also other linear equilibria, including the ex-ante worst linear score: a score may be state-wise optimal yet ex ante the worst linear one. The key to \cref{prop:normal} is the equivalence between equilibrium scores and the stationary points of the ex-ante payoff maximization problem over linear scores, which correspond to the eigenvectors of a matrix determined by the preference parameters and the correlation structure.

We give conditions under which the weights on two dimensions of the state in the best and worst equilibrium linear scores have the same sign, for a two-dimensional state (\cref{prop:normal_n=2}) and for an arbitrary number of dimensions (\cref{prop:normal_n}). In \cref{prop:comparative}, we analyze how welfare varies with the variance and covariance in the best and worst linear equilibria. A special case is when the state has two dimensions and the players want to minimize a weighted sum of the posterior variances. Then, there are two equilibrium linear scores (up to payoff equivalence). One score positively correlates actions across dimensions, the other negatively correlates them. The optimality of each score depends on the correlation structure: when the dimensions are positively correlated, the ex-ante best linear score correlates the actions, and vice versa. Accordingly, an increase in the absolute correlation is good for welfare in the best equilibrium but also increases the cost of coordinating on the wrong equilibrium. An increase in variance is unambiguously bad in any equilibrium.

\subsection{Related Literature}

We introduce a notion of equilibrium simple language by requiring the sender to use an aggregator of a multidimensional state, which we dub score. Closest to our paper is the literature studying cheap talk models with aligned preferences and some form of language limitation.\footnote{This literature, like us, looks at the consequences of language limitations, not its causes. On the latter topic see \citet{lipman2003} and \citet{lipman2025}.} \citet{jaeger_et_al2011} consider a  sender constrained to use a finite number of messages. They characterize the best equilibria and study their stability. 
In \citet{blume_board2013} and \citet{blume2018} uncertainty about the language used can impede communication. These papers model situations in which simplicity is a constraint on the number of messages and therefore on the set of strategies. As a consequence, they find that ex-ante optimal strategies are equilibrium strategies. Similarly, \citet{lipman2025} uses the fact that optimal strategies are equilibrium strategies to show that there is always an equilibrium in pure strategy when preferences are aligned in cheap-talk games with a possibly constrained set of messages. We take a different approach to modeling simple language: we introduce the notion of scores and require simplicity to be an equilibrium property instead of a constraint on the set of strategies itself. In particular, the optimal scores are not necessarily equilibrium strategies and therefore strategic frictions impose constraints on communication beyond the properties of scores.\footnote{Other papers consider scores that aggregate a multidimensional variable in different settings. For example, \citet{ball2025} and \citet{bonatti_cisternas2020} study linear scores where sender and receiver have different payoffs and the sender can manipulate the score input.}
 
We also relate to the literature on multidimensional cheap talk. This literature has shown that multiple dimensions can be useful for information revelation, e.g., \citet{battaglini2002}, \citet{chakraborty_harbaugh2007} and \citet{chakraborty_harbaugh2010}. In this strand of the literature, the contributions closest to ours are \citet{levy_razin2007} and \citet{lipnowski_ravid2026}. \citet{levy_razin2007} show that correlation across dimensions can limit communication   
by creating informational spillovers across dimensions. Similar mechanisms are at play in our paper as the sender needs to balance how the score, a one-dimensional object, reveals information across both dimensions. \citet{lipnowski_ravid2026} study a model where the dimension of the message is lower than the dimension of the state and players' strategies must be linear. Unlike us, they allow the dimension of the message to be larger than one and study how the sender groups issues across the different dimensions of the message. Instead, we do not assume linearity and show that dimension reduction and the IVP result in equilibria where strategies are quasilinear in the state.\footnote{In \cref{sec:normal} where we assume a normally distributed state, our characterization of linear equilibrium scores essentially coincides with theirs for the case where they assume a one-dimensional message.}

Finally, there is a strand of the literature in information design, i.e., communication with commitment, where the amount of information transmitted is limited. In \cite{gentzkow_kamenica2014}, the limitation comes from the cost of designing the experiment, while in  \cite{bloedel_segal2021} it comes from the information-processing cost faced by the receiver. When considering the commitment problem, we impose a restriction directly on the shape of the information structure, by limiting the sender to select among scores. In this way we are closer to 
\cite{letreust_tomala2019} and \cite{aybas_turkel2024}, who consider exogenous constraints on the capacity or cardinality of the message space.

\section{Model}

\subsection{Setting}\label{sect:setting}

\paragraph{Players, States and Strategies.} There are two players: a sender and a receiver. The sender has private information about an $n$-dimensional state of the world, $\theta=(\theta_1,..,\theta_n)\in\Theta = \reals^n$, with $n\geq 2$, whose distribution has full support and admits a density function $f$. When there is risk of ambiguity, we write $\Tilde{\theta}$ for the random variable and $\theta$ for its realization. Without loss of generality, we assume that $\E[\theta]=0$. 
We also assume that the variance of $\theta$ is finite. The receiver takes an $n$-dimensional action  $a=(a_1,..,a_n)\in\reals^n$. Before the receiver   acts, the sender sends a cheap-talk message $m\in \mathcal{M}$ for some measurable space $\mathcal{M}\supseteq \reals$.\footnote{We endow subsets of $\reals^n$ with their Euclidean topology and Borel $\sigma$-algebra.}

\paragraph{Payoffs.} 
Sender and receiver share the payoff function 
\[u(a,\theta)=-(a-\theta)^\top\Phi(a-\theta)\] 
for some symmetric positive definite matrix $\Phi$. In \Cref{subsec:existencecommitment}, we prove equilibrium properties for a larger set of concave payoff functions.

  The sender and receiver's strategies are measurable functions denoted by $\mu:\Theta\rightarrow \mathcal{M}$ and  $\alpha:\mathcal{M} \rightarrow \reals^n$, respectively. For any $m\in \mathcal{M}$ and $i=1,..,n$, let $\alpha_i(m)$ denote the $i$-th component of $\alpha(m)$. 
 
\paragraph{Equilibrium Selection.} We study Perfect Bayesian Equilibria in which the sender's strategy takes the form of a score.

A \textsl{score} $s$ is a non-constant measurable function $s:\Theta\rightarrow\reals$ that satisfies the following property:\footnote{Requiring scores to be non-constant simplifies the exposition of our main results at no loss of generality.}
\begin{itemize}
\item[] Intermediate Value Property (IVP): for any $\theta,\theta'\in\Theta$ such that $s(\theta)>s(\theta')$, any $m\in [s(\theta'),s(\theta)]\cap s(\Theta)$ and any continuous function $\gamma:[0,1]\rightarrow\Theta$ with $\gamma(0)=\theta'$ and $\gamma(1)=\theta$, there is a $t\in [0,1]$ such that $s(\gamma(t))=m$.
\end{itemize}

We denote the collection of all scores by $\mathcal{S}$.  Scores  aggregate  a complex state, represented by $\reals^n$, into a simpler message, represented by a real number. The Intermediate Value Property requires that along any continuous path through the state space, the score attains every message $m$ used elsewhere ($m\in  s(\Theta)$) and lying between the images of the
path's endpoints ($m\in [s(\theta'),s(\theta)]$). 

The need for aggregation fits at least two scenarios. First, the receiver may lack the time or capacity to process a rich description of the state, as when a policymaker relies on a broad assessment rather than a detailed report. Second, a single message may need to serve multiple audiences, as when a public health expert gives public advice that cannot be tailored to individual circumstances.  We discuss scores in more detail in the next section.

    \begin{definition}
    A score $s$ is an equilibrium score if there exists a Perfect Bayesian Equilibrium  in which the sender's strategy is $s$, i.e.,  there is a receiver's strategy $\alpha$ such that \begin{itemize}
        \item for all $\theta\in \Theta$, and $m,m'\in s(\Theta)$,\begin{equation}\label{eq:SIC}
            s(\theta)=m\ \Rightarrow\ u\bigl(\alpha(m),\theta\bigr)\geq u\bigl(\alpha(m'),\theta\bigr),\tag{S-IC}
        \end{equation}
        \item for all $m\in s(\Theta)$,\begin{equation}\label{eq:RBR}
            \alpha(m)\in\arg\max_{a\in\reals^{n}}\ \E\bigl[u(a,\tilde\theta)\bigm\vert s(\theta)=m\bigr].\tag{R-BR}
        \end{equation}
    \end{itemize}
\end{definition} 

In equilibrium, the sender in each state prefers the message assigned by the score over any other message given the receiver’s strategy (\ref{eq:SIC}), while the receiver best-replies given his beliefs, which are induced by the equilibrium score (\ref{eq:RBR}). We can ignore off-path messages in the definition of equilibrium score as any off-path message can be assigned a belief corresponding to an on-path message.\footnote{The definition implicitly assumes the existence of a conditional expectation. By the Disintegration Theorem \citep[e.g.,][Theorem 3.4]{kallenberg_2021}, a regular conditional distribution exists and therefore a conditional expectation exists everywhere.}

We refer to a PBE in which the sender's strategy is a score and induces at least two distinct posteriors simply as an \textsl{equilibrium}.
We call an equilibrium score a \textsl{best equilibrium score} if it maximizes the players' payoffs among all equilibrium scores.

We select  equilibria in which the sender's strategy is a score. This is different from constraining the sender to use only scores.\footnote{Examples of cheap-talk models with restrictions that, unlike ours, apply directly to the sender's set of strategies include \citet{jaeger_et_al2011}, \citet{blume_board2013}, \citet{lipman2025} and \citet{GordonKartikLoOlszewskiSobel2026}.} In our approach, off path the sender can select any strategy. This fits the scenarios described above, where the sender is an expert who can handle the complexity of the state and can, in principle, communicate it in any way. In particular, once the state is realized, the sender has the flexibility to send any message.

\subsection{Discussion of Scores}\label{sect:score}

Scores capture several desirable properties of simple communication on complex multidimensional issues. Scores require that messages have a lower dimension than states and that similar states must result in similar messages. 

Scores impose natural constraints, they are flexible, and they prevent injective mappings. Scores can be continuous or discrete mappings from $\reals^n$ to $\reals$. When the set $[s(\theta'),s(\theta)]\cap s(\Theta)$ is an interval, the IVP guarantees the familiar Intermediate Value Theorem holds along every curve connecting  $\theta$ and $\theta'$. When the set $[s(\theta'),s(\theta)]\cap s(\Theta)$ is discrete, the IVP requires that when moving from $\theta'$ to $\theta$, the score does not ``jump'' over any message. As an illustration of the latter case, consider an economist using coarse categories such as ``small'', ``moderate'' and ``large'' to describe the size of an economic shock. 
The property ensures that a marginal change in the state moves the score by at most one category. 
The IVP, a weak form of continuity, incorporates the natural convention that similar states must result in similar messages.\footnote{This convention might not fit all communication settings, but we think it is reasonable when the state has natural topological properties, as when assessing the size of a macroeconomic shock.}

Scores can take many forms. All continuous mappings from $\reals^n$ to $\reals$ are scores, but scores can also be discrete. They also need not be monotonic.  
Examples of scores include:
\begin{itemize}
    \item $s(\theta)=\beta^\top\theta$,
    \item $
    s(\theta)=\begin{cases}
        1&\text{if }\beta^\top\theta\geq k\\
        0&\text{otherwise}
    \end{cases}$,
    \item $s(\theta)=\max_{i\leq n}\theta_i$,
      \item $s(\theta)=\|\theta-\theta_0\|$ for some $\theta_0\in\reals^{n}$.
\end{itemize}

Scores prevent injective mappings from $\Theta$ into  the message space.\footnote{To see this, take three messages $m,m_1,m_2\in s(\Theta)$ with $m_1<m<m_2$, and states $\theta,\theta^1,\theta^2$ such that $s(\theta)=m$, $s(\theta^1)=m_1$ and $s(\theta^2)=m_2$. We can draw a curve from $\theta^1$ to $\theta^2$  that does not pass through $\theta$. By the IVP, there must be a point $\theta'''$ on that curve such that $s(\theta''')=m$.} By preventing injections, scores  prevent the sender from fully revealing the state and guarantee that the communication strategy truly reduces the dimension of the state space. 
Just requiring that the equilibrium strategy is a mapping from $\Theta$ to $\reals$ would not be enough: injections from $\reals^n$ to $\reals$ exist. To rule out injections,  we additionally require strategies to satisfy the IVP.\footnote{Our approach parallels the one adopted in the literature on informational requirements in mechanism design. This literature imposes continuity-like assumptions on mechanisms to avoid bijections between sets of different dimensions. See for example, \citet{hurwicz1972}, \citet{mount_reiter1974} and \citet{reichelstein1984}.}
 
\section{Illustration in a 2x2 Setting}\label{sec:2X2}

A consequence of our equilibrium selection is that the payoffs from committing to a score can exceed any equilibrium payoffs, despite the common interest between sender and receiver. Here we illustrate this point in a  2x2 setting. The analysis in this section is self-contained.

Abusing notation slightly, let $\Theta=\{0,1\}^2$ and $f(\theta)$ denote the probability of state $\theta\in\{0,1\}^2$.  We focus on  symmetric quadratic payoffs, i.e., we set $\Phi$ equal to the $2\times2$ identity matrix, and we fix the prior  $f(\theta)=\frac{1-\epsilon}{3}$ for $\theta\neq (1,0)$ and $f(1,0)= \epsilon $ for some $\epsilon\in(0,\frac{3}{\sqrt{2}}-2)$, so that $(1,0)$ is considerably less likely than the other states.

Our definition of scores does not apply directly to this finite environment but it can be adapted by requiring that the score does not ``jump'' messages when moving through points that differ by at most one coordinate. Formally, this can be done by replacing the topological notion of path (the function $\gamma$ in the definition) with the graph-theoretic notion of path. Viewing the square $\{0,1\}^2$ as a simple graph where each point is a vertex and the edges of the graph are the edges of the square, we can rewrite the IVP as follows: for any $\theta,\theta'\in\Theta$ with $s(\theta)>s(\theta')$, any $m\in [s(\theta'),s(\theta)]\cap s(\Theta)$ and any path $p$ from $\theta'$ to $\theta$, there is some $\theta''$ on the path such that $s(\theta'')=m$.\footnote{See e.g., \citet{west2001} for definitions.}

Let score $s_d$  be as shown in the left panel of  \cref{fig:valcomm}. The score $s_d$ assigns a common message  to states $(0,1)$ and $(1,0)$ while assigning unique messages  to the other states. We show in the appendix that, in the setting we consider, score $s_d$ ensures a higher expected payoff than any other score (up to a relabeling of the messages). In other words, if the sender could commit, she would pick score  $s_d$. At the same time, $s_d$ is not an equilibrium score. The intuition is as follows. As by assumption  $ f(0,1)$ is sizably larger than $f(1,0)$,   the  posterior associated with $m=2$ lies ``close'' to $(0,1)$ and ``far'' from $(1,0)$. In fact, the posterior is so far from $(1,0)$ that  if the receiver expects the sender to communicate according to the score, i.e., $\mu(\theta)=s_d(\theta)$ for all $\theta$, then the sender has a profitable deviation upon observing state $(1,0)$. The right-hand side of \cref{fig:valcomm} shows the resulting deviating strategy at the ex-ante stage, which involves message $\mu(1,0)=3$ instead of $\mu(1,0)=2$.

\begin{figure}[h]
\begin{center}
\begin{subfigure}{.3\textwidth}
\centering
\begin{tikzpicture}[scale=2]
    \fill (0,0) circle (2pt);
    \fill (1,0) circle (2pt);
    \fill (0,1) circle (2pt);
    \fill (1,1) circle (2pt);
  \fill[fill= lightgray, opacity=0.5,rotate=45] (.7,0) ellipse (.3 and .9);
   \draw[  opacity=0.5] (0,0) ellipse (.25 and .25);
     \fill[fill=darkgray,opacity=0.5] (1,1) ellipse (.25 and .25);

  \node[mark size=4pt, thick, color=black] at (0.15,0.85) {\pgfuseplotmark{x}};
\end{tikzpicture}
\caption{Strategy $\mu=s_d$.}
\end{subfigure}
\begin{subfigure}{.3\textwidth}
\centering
\begin{tikzpicture}[scale=2]
    \fill (0,0) circle (2pt);
    \fill (1,0) circle (2pt);
    \fill (0,1) circle (2pt);
    \fill (1,1) circle (2pt);
     \fill[ fill=darkgray,opacity=0.5] (1,.5) ellipse (0.3 and .8);
     \fill[fill= lightgray, opacity=0.5,] (0,1) ellipse (.25 and .25);
     \draw (0,0) ellipse (.25 and .25);
\end{tikzpicture} 
\caption{Profitable deviation}
\end{subfigure}
\begin{subfigure}{.1\textwidth}
\hspace{1cm}
\vspace{1cm}
\begin{tikzpicture}
    \draw[thick] (-0.2, -1.95) rectangle (2.5, 0.4);
    \fill[gray]  (0,0) rectangle (0.2,0.2);
    \node[right] at (0.3,0.1) {$m=3$};
    \fill[lightgray] (0,-0.5) rectangle (0.2,-0.3);
    \node[right] at (0.3,-0.4) {$m=2$};
    \draw (0,-1) rectangle (0.2,-0.8);
    \node[right] at (0.3,-0.9) {$m=1$};
    \node[mark size=4pt, thick, color=black] at (0.1,-1.6) {\pgfuseplotmark{x}};
    \node[right] at (0.3,-1.6) {$\E[\theta\vert m=2]$};
\end{tikzpicture}
\end{subfigure}
\end{center}
\caption{}\label{fig:valcomm}\end{figure}

The deviating strategy violates the IVP, as it ``jumps'' from $\mu(0,0)=1$ directly to $\mu(1,0)=3$.  This strategy is  not a score. In general,   scores that maximize payoffs need not be equilibrium strategies precisely because deviations to strategies that are not scores are possible. Relatedly, in some cases, the players 
are better off if the sender  observes only one dimension of the state  (see \Cref{app:ignorance}). The intuition is that ignorance reduces the set of potential deviations available to the sender.  This is in contrast with the rest of the literature that studies cheap talk models with aligned preferences (\citet{jaeger_et_al2011}, \citet{blume_board2013} and \citet{blume2018}) where the constraints on communication bind the message space directly.

\section{Analysis}

In \Cref{subsec:existencecommitment}, we establish that equilibrium scores exist and that equilibrium payoffs lie strictly below the commitment payoffs; these results hold for a large class of concave payoff functions. In the remainder of the section, we return to quadratic payoffs. \Cref{sec:Analysis} shows that equilibrium scores are payoff-equivalent to scores that are monotonic transformations of a linear function. In \Cref{sec:normal}, we additionally assume that states are normally distributed and focus on scores linear in the state.

\subsection{Equilibrium Properties}\label{subsec:existencecommitment}

The results in this subsection hold for a large class of convex loss functions (see \Cref{app:proofProp:existence}). We first establish existence of an equilibrium score.
\begin{proposition}\label{prop:existence}
    An equilibrium exists.
\end{proposition}
 We establish existence by showing that there always exists a PBE with two messages in the support of the sender's strategy.\footnote{Here we build on  the proof of existence in \citet{jaeger_et_al2011}. We adapt the proof to accommodate an unbounded state space and sender deviations to strategies that are not scores.} Since any   strategy of the sender with two messages trivially satisfies   the properties of a score, an equilibrium score exists.

The example in \cref{sec:2X2} shows that, despite the alignment of preferences between sender and receiver, the 
incentive compatibility constraints (\ref{eq:SIC}) limit the information transmission beyond the restriction imposed by scores. 
\cref{proposition:calendar} generalizes this insight. We make this notion precise by considering a sender who can commit to a score, i.e. need not satisfy the incentive-compatibility constraints.
Let $v^*$ denote the \textsl{commitment payoff}, i.e.,  
\[
    v^{*}:=\sup_{s\in\mathcal S}\ \E\bigl[u(\alpha(s(\theta)),\theta)\bigr]
    \text{ \quad s.t. \quad }
    \alpha(m)\in\arg\max_{a\in\reals^{n}}
    \E\bigl[u(a,\tilde\theta)\mid s(\theta)=m\bigr]
    \ \forall m\in s(\Theta).
\]
\begin{proposition}
\label{proposition:calendar}
    Equilibrium payoffs are uniformly below the commitment payoff.
\end{proposition}

In \cref{sec:2X2}, we show in an example that commitment has value by constructing a score that yields strictly higher payoffs than any equilibrium score. For the general case, we instead construct sequences of scores whose payoffs converge to $0$, the payoff under full disclosure. Since payoffs cannot exceed $0$, the commitment payoff is $0$. The scores in these sequences are not equilibrium scores, as they violate the sender’s incentive-compatibility constraints \eqref{eq:SIC}. To conclude the proof of \cref{proposition:calendar}, we characterize a uniform bound, strictly below $0$, such that payoffs in all equilibria lie below it.

Broadly, the proof of \cref{proposition:calendar} goes as follows. Scores are not injections. This implies that, for any equilibrium score, some states must be bounded away from their assigned action. By \ref{eq:SIC},  states  bounded away from their assigned actions must be bounded away from all induced actions in equilibrium. Therefore there is a non-vanishing region around these states where the receiver does not match his action with the state. This argument establishes that equilibrium payoffs are bounded away from $0$. Instead, when the sender does not need to satisfy incentive-compatibility constraints, the regions where the receiver does not match his action with the state can be made vanishingly small by constructing an appropriate sequence of scores. Therefore, $v^*=0$.

The result that $v^*=0$ shows that the definition of scores is flexible and in a sense, minimally restricts the mappings from $\reals^n$ to $\reals$. It also shows that, if one were interested in the commitment problem of the sender, the IVP does not have enough bite and one should arguably impose stronger restrictions like a form of Lipschitz continuity. Since the commitment problem is not our focus, we allow for the more permissive IVP.  The result, however, offers a clean benchmark against which to compare equilibrium scores. We quantify  the loss from the lack of commitment in \cref{sec:normal}, where we assume normally distributed states.

\subsection{Equilibrium Scores}\label{sec:Analysis}

In this subsection and the next we focus on the case where the loss function takes the quadratic form defined in \Cref{sect:setting}. We show that equilibrium scores must satisfy specific properties.   We characterize these properties for arbitrary priors in this subsection, and specialize to normally distributed states in the next. 

Two equilibrium scores $s$ and $s'$ are \textsl{payoff-equivalent} if the equilibria they induce give players the same expected payoff.

\begin{proposition}\label{prop:linear}
    For any equilibrium score, there is a payoff-equivalent equilibrium score of the form $$
    s(\theta)=h(\beta^\top\theta),
    $$for some $\beta\in\reals^n\setminus\{0\}$ and increasing function $h:\reals\rightarrow\reals$ such that receiver's equilibrium action is $\alpha(m)=b+m\cdot \Phi^{-1}\beta$ for some $b\in\reals^n$.
\end{proposition}

\cref{prop:linear} records two consequences of the strategic frictions in our environment. First, it is without loss to focus on equilibrium scores that satisfy a specific functional-form restriction. Despite the wide range of admissible scores, including some that approach full revelation, in equilibrium, the sender uses a score that first aggregates the state linearly then potentially coarsens it further.\footnote{Were the loss function not quadratic, the equilibrium forces would induce alternative constraints on the admissible functional forms of the score. In light of this observation, \Cref{prop:linear} should not be interpreted as demonstrating that linearity is inherently special, but rather as illustrating that strategic frictions generate functional-form restrictions on the structure of communication.} Second, in any equilibrium, equilibrium actions must lie on a line. 

To illustrate the restrictions that $h(\beta^\top\theta)$ imposes, \cref{fig:eqac} plots two specifications of $h$ for $\Theta=\reals^2$. Panel (a) shows a case in which $h$ is strictly increasing. Each solid line collects the states that share a common value of $\beta^\top\theta$, distinct lines are assigned distinct scores, and the circles mark the equilibrium actions.   Panel (b) shows a case in which $h$ is a step function. Each band pools the states for which $\beta^\top\theta$ falls within a flat portion of $h$. Any score in which $h$ is strictly increasing is payoff-equivalent to one in which $h$ is the identity, i.e., a score that is linear in $\theta$. A step function instead coarsens the information the decision-maker receives.

\begin{figure}[h]
    \centering
  \begin{subfigure}{0.45\textwidth}
        \centering
\begin{tikzpicture}
  \clip (-1, -1) rectangle (5, 5);
  \draw[->] (-1, 0) -- (4.5, 0) node[right] {$\theta_1$};
  \draw[->] (0, -1) -- (0, 4.5) node[above] {$\theta_2$};
  \foreach \i in {0,...,8} {
    \pgfmathsetmacro{\offset}{1.7 + \i * (2.8 / 8)}
    \pgfmathsetmacro{\shade}{(\i+2) * 100 / 10}
    \draw[thick, black!\shade] (-1, {\offset}) -- (4.4, {\offset - 0.8*5.4});
  }
  \draw[thick, dashed] (-0.8, -1) -- (3.2, 4);
  \foreach \i in {0,...,8} {
    \pgfmathsetmacro{\offset}{1.7 + \i * (2.8 / 8)}
    \pgfmathsetmacro{\shade}{(\i+2) * 100 / 10}
    \pgfmathsetmacro{\xi}{(\offset - 0.8) / 2.05}
    \pgfmathsetmacro{\yi}{1.25*\xi}
    \filldraw[draw=black, fill=black!\shade] (\xi, \yi) circle (2pt);
  }
\end{tikzpicture}
        \caption{The function $h$ is strictly increasing.}
    \end{subfigure}
    \hfill
      \begin{subfigure}{0.45\textwidth}
        \centering
    \begin{tikzpicture}
  \clip (-1, -1) rectangle (5, 5);
  \draw[->] (-1, 0) -- (4.5, 0) node[right] {$\theta_1$};
  \draw[->] (0, -1) -- (0, 4.5) node[above] {$\theta_2$};
  \draw[thick, black] (-1, 2.5) -- (4.4, 2.5-0.8*5.4);
  \draw[thick, black] (-1, 4.4) -- (4.4, 4-0.8*5.4);
  \fill[opacity=0.3, black!10] (-1, 2.5) -- (-1,4.5) -- (4.4, {4-0.8*5.4}) -- (4.4, {2.5-0.8*5.4}) -- cycle;
  \fill[opacity=0.3, black!50] (-1, 4.4) -- (4.4,{4-0.8*5.4}) -- (4.4, 4.4) -- cycle;
  \draw[thick, dashed] (-0.8, -1) -- (3.2, 4);
  \filldraw[fill=white, draw=black]   (-0.6, -0.75) circle (2pt);
  \filldraw[fill=black!10, draw=black]    (1.25, 1.5625) circle (2pt);
  \filldraw[fill=black!50, draw=black] (2.4, 3) circle (2pt);
\end{tikzpicture}
        \caption{The function $h$ is a step function.}
    \end{subfigure}
    \caption{\label{fig:eqac} Equilibrium Scores and Actions}
\end{figure}

To see the logic behind \Cref{prop:linear}, observe that, given a belief about the sender's strategy, the receiver takes the action $\alpha(m)=\E[\theta\vert m]$. The sender problem in state $\theta$, given this belief, is to choose the message $m'$ that minimizes the loss function:$$
\min_{m'} \left((\theta-\alpha(m'))^\top\Phi(\theta-\alpha(m'))\right).
$$ 

Because the sender minimizes a weighted Euclidean distance, the set of states indifferent between any two messages is a hyperplane. The IVP requires that these indifference hyperplanes do not cross. This observation delivers the two main claims in \cref{prop:linear}. First, the equilibrium actions are collinear. Second, there is a unique slope $\beta$ such that if some type $\theta$ prefers to induce action $a$, then every type $\theta'$ with $\beta^\top\theta'=\beta^\top\theta$ also prefers to induce $a$.\footnote{There are two reasons we get the characterization up to payoff-equivalence. First, types with more than one optimal message, who have probability zero in equilibrium, could send different messages. Second, there could be equilibria where different messages induce the same action. Merging these messages would induce the same distribution over outcomes.}

\cref{prop:linear} establishes that we can without loss focus on equilibria where $$\alpha(m)=\E[\theta\vert m]=b+m\cdot\Phi^{-1}\beta$$ for some $b\in\reals^n$. The first equality is the receiver's best-reply. The second comes from the incentive compatibility constraints \eqref{eq:SIC}. Two observations follow. 

First, under any equilibrium score, the induced mapping from messages to the receiver’s actions is monotone in each dimension.\footnote{This is not an immediate consequence of linearity: a linear score need not induce a monotone mapping from messages to actions.} In particular, in equilibrium each dimension $i$ of the action falls into one of three categories, according to whether $(\Phi^{-1}\beta)_i$ is positive, zero, or negative. For any two dimensions in the same category, a higher message is associated with a change in the  expected state with the same sign. Across categories, the score can be interpreted as a relative measure: a higher message is associated with a higher expected value in some dimensions of the state and a lower expected value in others.

Second, scores that reveal a single dimension of the state (i.e., scores with $\beta_i=0$ for all but one dimension) are equilibrium scores only for a non-generic set of parameters. To see this point, suppose that $\Phi=\text{diag}(\phi_1,\dots,\phi_n)$ and that the sender uses a score that fully reveals one dimension, say $s(\theta)=\theta_1$. \Cref{prop:linear} implies that in equilibrium, $\E[\theta_i\vert m]=b_i+0\cdot\frac{m}{\phi_i}$ for all $i\neq 1$. In other words, because the score fully reveals $\theta_1$, the conditional expectation $\E[\theta_i\vert m=\theta_1]$ does not depend on $\theta_1$. 
The intuition is that if $\theta_1$ were informative about $\theta_i$, the sender would gain from deviating to correct the receiver’s expectations about $\theta_i$. This information spillover echoes \citet{levy_razin2007}, who show that misalignment in one dimension can hinder communication in another dimension where the sender and receiver have aligned preferences.

\subsection{Normally Distributed State}\label{sec:normal}
We now impose more structure on the prior by assuming that the state is normally distributed, and we fully characterize the linear equilibrium scores, including their welfare properties. The characterization of linear scores enables us to determine the best equilibrium score. 
The results of this subsection apply to elliptical distributions, a broader class of distributions that includes the multivariate normal, logistic, and  $t$ distributions.

In this section, we assume that $\theta\sim N(0,\Sigma)$ where $\Sigma$ is a covariance matrix. We call a score \textsl{linear} if it belongs to $\mathcal S_l:=\{s(\theta)=\beta^\top\theta:\beta\in\reals^n\setminus\{0\}\}$.
We refer to a score as an \textsl{ex-ante best linear score} if it solves the problem: 
\begin{align*}
   &\max_{s\in \mathcal{S}_l} \E[-(\alpha(s(\theta))-\theta)^\top\Phi(\alpha(s(\theta))-\theta)] \quad 
   \text{ s.t. }\alpha(m)=\E[\theta\vert m],\quad\forall m\in s(\Theta).
\end{align*}
Analogously, we refer to a score as an \textsl{ex-ante worst linear score} if it solves the problem with $\min$ instead of $\max$.

Ex-ante best and worst linear scores need not be in general equilibrium scores. The following proposition establishes that, for normally distributed states, they are.  

\begin{proposition}\label{prop:normal}
    Let $\theta\sim N(0,\Sigma)$. A linear score $s(\theta)=\beta^\top\theta$ is an equilibrium score if and only if $\beta$ is an eigenvector of $\Phi\Sigma$. The set of equilibrium linear scores include the ex-ante best and worst linear scores.
\end{proposition}

\begin{corollary}\label{cor:best}
    Let $\theta\sim N(0,\Sigma)$. The best equilibrium score is an ex-ante best linear score.
\end{corollary}

From now on, we identify a linear score $s(\theta)=\beta^\top\theta$ with the weights $\beta$.

 \Cref{prop:normal} shows that, for normally distributed states, there exists an equilibrium that induces the ex-ante best linear score. Other  equilibria exist too, including one that induces the ex-ante worst linear score.
The key idea underlying the proof is the following.  
A linear score $\beta$, with  corresponding strategy $\alpha$, is an equilibrium score if the indifference curve of each type $\theta$ sending message $m$,  $\{a\in \reals^2:u(a,\theta)=u(\alpha(m),\theta)\}$, is tangent to the curve $\{\alpha(m):m\in\reals\}$. We show that the linear scores $\beta$ satisfying these tangency conditions are the eigenvectors of $\Phi\Sigma$. These eigenvectors, in turn, solve the first-order conditions of the ex-ante maximization problem.

\cref{cor:best} follows from \cref{prop:linear} and \cref{prop:normal}. Take any equilibrium score $s$. By \cref{prop:linear}, there is a payoff-equivalent equilibrium score $h(\beta^\top\theta)$ for some $\beta\neq0$ and increasing $h$. Since $h(\beta^\top\theta)$ is a function of $\beta^\top\theta$, it conveys weakly less information than the linear score $\beta^\top\theta$, and, because the receiver best-replies to the information he receives, the players' payoff under $h(\beta^\top\theta)$ is weakly lower than under $\beta^\top\theta$. Hence the payoff from $s$ is weakly below the payoff from the linear score $\beta^\top\theta$, which in turn is weakly below the payoff from an ex-ante best linear score. By \cref{prop:normal} an ex-ante best linear score is itself an equilibrium score, so it is a best equilibrium score.

\paragraph{The weights in ex-ante best and worst linear scores.} Using \cref{prop:normal}, we can characterize the equilibrium scores, in particular the best and worst equilibrium linear scores. Recall that from \cref{prop:normal}, these correspond to the ex-ante best and worst linear scores. We first focus on the case $n=2$. 

\begin{proposition}\label{prop:normal_n=2}
    Suppose $\theta\sim N(0,\Sigma)$ and $n=2$. Suppose also that $\Phi\Sigma\neq \lambda I$ for all $\lambda\in\reals$.

    Take a best and worst equilibrium linear score $\beta^+,\beta^-$. Then,\begin{align*}
        &\beta^+_1\beta^+_2\geq 0\Leftrightarrow \sigma_{12}\geq -\phi_{12}\max\left\{\frac{\sigma_1^2}{\phi_2},\frac{\sigma_2^2}{\phi_1}\right\},\\
        & \beta^-_1\beta^-_2\geq 0 \Leftrightarrow \sigma_{12}\leq -\phi_{12}\min\left\{\frac{\sigma_1^2}{\phi_2},\frac{\sigma_2^2}{\phi_1}\right\}.
    \end{align*}
\end{proposition}

\Cref{prop:normal_n=2} states when the weights on the two dimensions have the same sign in the best equilibrium linear score, and when they do in the worst. If $\Phi\Sigma$ is proportional to the identity matrix, then all linear scores are equilibrium scores and give the same payoffs. To understand these conditions better, first consider the case where $\Phi$ is diagonal, i.e., $\phi_{12}=0$. In that case, in the best equilibrium score, the weights $\beta_1$ and $\beta_2$ of the best equilibrium score have the same sign whenever the covariance is positive. Intuitively, a positive covariance, combined with weights of the same sign, positively correlates the actions of the receiver. Correlating the actions positively matches the state better when the state is positively correlated. Conversely, the worst equilibrium negatively correlates the receiver's actions when the state is positively correlated. The worst linear score can be a natural candidate for an equilibrium score. For example, if movie critics use a rating system in which a higher rating indicates higher aesthetic or entertainment value but these two dimensions are negatively correlated, then the equilibrium score has poor welfare properties.

To understand the role of $\phi_{12}$, consider  what happens when $\sigma_{12}=0$. In that case, if $\phi_{12}>0$, then the weights $\beta_1$ and $\beta_2$ of the best equilibrium linear score have  the same sign. Indeed, when the receiver best replies, the players' ex-ante utility is $-\E\big[\phi_1\text{Var}[\theta_1\vert m]+2\phi_{12}\text{Cov}[\theta_1,\theta_2\vert m]+\phi_2\text{Var}[\theta_2\vert m]\big]$. Therefore, with a positive $\phi_{12}$, the players want to minimize the covariance between the two dimensions of the state conditional on the message. Since $\theta_1=\frac{m}{\beta_1}-\frac{\beta_2}{\beta_1}\theta_2$, a high state in one dimension is associated with a low state in the other when both $\beta_1,\beta_2$ have the same sign. Therefore, the ex-ante best linear score has $\beta_1\beta_2\geq 0$ and the worst one has $\beta_1\beta_2\leq 0$.

The results for $n=2$ can be partially generalized to arbitrary dimensions. 

\begin{proposition}\label{prop:normal_n}
    Suppose $\theta\sim N(0,\Sigma)$ and fix some $i,j\in\{1,...,n\}$, $i\neq j$. Let $\Sigma,\hat{\Sigma}$ be two covariance matrices such that $\sigma_{ij}<\hat{\sigma}_{ij}$ and $\Sigma_{k\ell}=\hat{\Sigma}_{k\ell}$ for all $(k,\ell)\neq (i,j)$.

    Then, for any best and worst equilibrium linear scores under $\Sigma$, $\beta^+,\beta^-$, and under $\hat{\Sigma}$, $\hat{\beta}^+,\hat{\beta}^-$, we have \begin{align*}
        &\beta^+_i\beta^+_j\geq 0 \quad\Rightarrow\quad  \hat{\beta}^+_i\hat{\beta}^+_j\geq 0,\\
        \text{and }\;&\beta^-_i\beta^-_j\leq 0 \quad \Rightarrow \quad \hat{\beta}^-_i\hat{\beta}^-_j\leq 0.
    \end{align*}
\end{proposition}

For arbitrary dimensions of the state space, \Cref{prop:normal_n} shows that the sign of $\beta_i\beta_j$ has a single-crossing structure similar to that in \cref{prop:normal_n=2}. In the ex ante best linear score, $\beta_i$ and $\beta_j$ have (weakly) the same sign when $\sigma_{ij}$ is sufficiently large.\footnote{The covariance threshold above which $\beta_i\beta_j\geq 0$ may be unattainable.} In contrast, in the ex ante worst linear score, $\beta_i$ and $\beta_j$ have opposite signs when $\sigma_{ij}$ is sufficiently large.

\paragraph{Welfare analysis.} We now characterize how payoffs in the best and worst linear equilibrium vary with the covariance matrix.

\begin{proposition}\label{prop:comparative}
    Suppose $\theta\sim N(0,\Sigma)$. Let $v^+(\Sigma)$ and $v^-(\Sigma)$ be the expected utility from the best and worst equilibrium linear scores. For every $i$ and $j$:\begin{itemize}
        \item The values $v^+(\Sigma)$ and $v^-(\Sigma)$ are  decreasing in $\sigma^2_i$.
        \item The value $v^+(\Sigma)$ is convex in $\sigma_{ij}$.\begin{itemize}
            \item If $n=2$, a minimum is attained at $\sigma_{12}=-\phi_{12}\min\left\{\frac{\sigma_1^2}{\phi_2},\frac{\sigma_2^2}{\phi_1}\right\}$.
        \end{itemize}
        \item The value $v^-(\Sigma)$ is concave in $\sigma_{ij}$.\begin{itemize}
            \item If $n=2$, a maximum is attained at $\sigma_{12}=-\phi_{12}\max\left\{\frac{\sigma_1^2}{\phi_2},\frac{\sigma_2^2}{\phi_1}\right\}$.
        \end{itemize}
        \end{itemize} 
\end{proposition}

\Cref{prop:comparative} provides comparative statics for the players’ welfare with respect to the covariance matrix. First, it establishes that an increase in a variance lowers the players’ utility in both the best and the worst linear equilibrium. Second, it shows that an increase in the covariance has an ambiguous effect on utility. In the best linear equilibrium, a higher covariance first decreases and then increases payoffs; in the worst linear equilibrium, it first increases and then decreases them. 

For $n=2$ and $\phi_{12}=0$, the result is equivalent to establishing that the best equilibrium payoff is increasing in $\vert\sigma_{12}\vert$ and the worst equilibrium payoff is decreasing in $\vert\sigma_{12}\vert$. Intuitively, higher covariance brings the two dimensions closer to one dimension. In this case, the best linear score captures the state well, but the worst score pools highly probable states together. With $\phi_{12}\neq 0$, the minimum of $v^+(\Sigma)$ is attained but the maximum of $v^-(\Sigma)$ may not be attained, since $\vert\sigma_{12}\vert\leq \sigma_1\sigma_2$.

\section{Conclusion}

We study simple communication in a cheap-talk game with aligned preferences. We model simplicity by requiring that in equilibrium the sender strategy is a score. This equilibrium selection requirement introduces demand for commitment despite the alignment of preferences between sender and receiver. The requirement also imposes structure on the functional form of equilibrium scores.

Our notion of simplicity combines dimension reduction and a weak form of continuity (the IVP). The multidimensionality of our model plays a central role and the IVP alone would not deliver the result. Indeed, if the state were one-dimensional, then, holding fixed  the quadratic-loss payoffs, a best equilibrium score would yield the commitment payoffs. In a one-dimensional setting, a score can be defined in several ways. Let $\Theta\subseteq \reals$ and let the sender send messages in $M\subseteq\reals$. A score is a function $s$  satisfying\begin{enumerate}
    \item $s:\Theta\rightarrow M$ and 
    \item $s$ satisfies the IVP. 
\end{enumerate}
If $M=\reals$, full revelation is possible, so the best equilibrium yields the commitment payoffs. If $M=\{1,...,n\}$, the equivalence between best equilibrium and commitment payoffs follows from the observation that for any given score and its associated belief, the most profitable deviation is itself a score. Hence, if such a deviation were profitable, the original score could not have attained the commitment payoffs in the first place. This is the key difference with the multidimensional case, in which a profitable deviation may be a strategy that is not a score.

\newpage
 
\bibliographystyle{agsm}
\bibliography{bib.bib} 

@article{battaglini2002,
author = {Marco Battaglini},
 journal = {Econometrica},
 number = {4},
 pages = {1379--1401},
 title = {Multiple Referrals and Multidimensional Cheap Talk},
 urldate = {2025-03-11},
 volume = {70},
 year = {2002}
}

@book{kallenberg_2021,
year = {2021},
author = {Kallenberg, Olav},
address = {Cham, Switzerland},
edition = {Third edition.},
isbn = {3-030-61871-4},
language = {eng},
publisher = {Springer},
series = {Probability Theory and Stochastic Modelling, 99},
title = {Foundations of modern probability},
}

@article{lipnowski_ravid2026,
    author = {Lipnowski, Elliot and Ravid, Doron},
    title = {Communicating about Endogenous Issues},
    year = {2026}
}

@book{casella2002statistical,
  author    = {George Casella and Roger L. Berger},
  title     = {Statistical Inference},
  edition   = {2nd},
  publisher = {Thomson Learning},
  year      = {2002},
  isbn      = {978-0534243128}
}

@article{gentzkow_kamenica2014,
  title={Costly persuasion},
  author={Gentzkow, Matthew and Kamenica, Emir},
  journal={American Economic Review},
  volume={104},
  number={5},
  pages={457--462},
  year={2014},
  publisher={American Economic Association 2014 Broadway, Suite 305, Nashville, TN 37203}
}

@article{hurwicz1972,
  title={On informationally decentralized systems},
  author={Hurwicz, Leonid},
  journal={Decision and organization: A volume in Honor of J. Marschak},
  year={1972},
  publisher={North-Holland}
}

@article{mount_reiter1974,
number = {2},
pages = {161-192},
publisher = {Elsevier Inc},
title = {The informational size of message spaces},
volume = {8},
year = {1974},
author = {Mount, Kenneth and Reiter, Stanley},
address = {New York},
copyright = {1974},
issn = {0022-0531},
journal = {Journal of Economic Theory},
}

@book{west2001,
  title={Introduction to graph theory},
  author={West, Douglas Brent},
  volume={2},
  year={2001},
  publisher={Prentice hall Upper Saddle River}
}

@article{reichelstein1984,
  title={Incentive compatibility and informational requirements},
  author={Reichelstein, Stefan},
  journal={Journal of Economic Theory},
  volume={34},
  number={1},
  pages={32--51},
  year={1984},
  publisher={Elsevier}
}

@article{lipman2025,
  title={Why is language vague?},
  author={Lipman, Barton L},
  journal={International Journal of Game Theory},
  volume={54},
  number={1},
  pages={8},
  year={2025},
  publisher={Springer}
}

@article{blume_board2013,
author = {Blume, Andreas and Board, Oliver},
title = {Language Barriers},
journal = {Econometrica},
volume = {81},
number = {2},
pages = {781-812},
year = {2013}
}

@article{bloedel_segal2021,
  title={Persuasion with rational inattention},
  author={Bloedel, Alexander W and Segal, Ilya R},
  journal={Available at SSRN 3164033},
  year={2021}
}

@article{blume2018,
author = {Blume, Andreas},
title = {Failure of common knowledge of language in common-interest communication games},
journal = {Games and Economic Behavior},
volume = {109},
pages = {132-155},
year = {2018}
}

@article{letreust_tomala2019,
  title={Persuasion with limited communication capacity},
  author={Le Treust, Ma{\"e}l and Tomala, Tristan},
  journal={Journal of Economic Theory},
  volume={184},
  pages={104940},
  year={2019},
  publisher={Elsevier}
}

@article{aybas_turkel2024,
  title={Persuasion with coarse communication},
  author={Aybas, Yunus C and Turkel, Eray},
  journal={arXiv preprint arXiv:1910.13547},
  year={2024}
}

@incollection{lipman2003,
  title={Language and economics},
  author={Lipman, Barton L},
  booktitle={Cognitive Processes and Economic Behaviour},
  pages={75--93},
  year={2003},
  publisher={Routledge}
}

@article{levy_razin2007,
  title={On the limits of communication in multidimensional cheap talk: a comment},
  author={Levy, Gilat and Razin, Ronny},
  journal={Econometrica},
  volume={75},
  number={3},
  pages={885--893},
  year={2007},
  publisher={Wiley Online Library}
}

@article{chakraborty_harbaugh2007,
  title={Comparative cheap talk},
  author={Chakraborty, Archishman and Harbaugh, Rick},
  journal={Journal of Economic Theory},
  volume={132},
  number={1},
  pages={70--94},
  year={2007},
  publisher={Elsevier}
}

@article{chakraborty_harbaugh2010,
  title={Persuasion by cheap talk},
  author={Chakraborty, Archishman and Harbaugh, Rick},
  journal={American Economic Review},
  volume={100},
  number={5},
  pages={2361--2382},
  year={2010},
  publisher={American Economic Association}
}

@article{jaeger_et_al2011,
title = {Voronoi languages: Equilibria in cheap-talk games with high-dimensional types and few signals},
journal = {Games and Economic Behavior},
volume = {73},
number = {2},
pages = {517-537},
year = {2011},
author = {Gerhard Jäger and Lars P. Metzger and Frank Riedel}}

@book{parlett1998,
  title={The symmetric eigenvalue problem},
  author={Parlett, Beresford N},
  year={1998},
  publisher={SIAM}
}

@article{lang1986,
  title={A note on the measurability of convex sets},
  author={Lang, Robert},
  journal={Archiv der Mathematik},
  volume={47},
  pages={90--92},
  year={1986},
  publisher={Springer}
}

@article{milgrom_segal2002,
author = {Milgrom, Paul and Segal, Ilya},
title = {Envelope Theorems for Arbitrary Choice Sets},
journal = {Econometrica},
volume = {70},
number = {2},
pages = {583-601},
year = {2002}
}

@article{bonatti_cisternas2020,
  title={Consumer scores and price discrimination},
  author={Bonatti, Alessandro and Cisternas, Gonzalo},
  journal={The Review of Economic Studies},
  volume={87},
  number={2},
  pages={750--791},
  year={2020},
  publisher={Oxford University Press}
}

@article{ball2025,
  title={Scoring strategic agents},
  author={Ball, Ian},
  journal={American Economic Journal: Microeconomics},
  volume={17},
  number={1},
  pages={97--129},
  year={2025},
  publisher={American Economic Association 2014 Broadway, Suite 305, Nashville, TN 37203-2425}
}

@unpublished{GordonKartikLoOlszewskiSobel2026,
  author = {Gordon, Sidartha and Kartik, Navin and Lo, Melody and Olszewski, Wojciech and Sobel, Joel},
  title  = {Effective Communication in Cheap-Talk Games},
  note   = {Working paper},
  year   = {2026},
  month  = mar
}

\newpage

\appendix
\renewcommand{\theequation}{A.\arabic{equation}}
\setcounter{equation}{0} 
\setcounter{lemma}{0}
\renewcommand{\thelemma}{A\arabic{lemma}}

\section{Proofs for \Cref{sec:2X2}}

In the text of \Cref{sec:2X2} we claim that, for the prior we focus on,  score $s_d$ ensures higher payoffs than any other score, and that $s_d$ is  not an equilibrium score. The following remark supports the claim.  We call a score \textsl{optimal} if it maximizes the payoffs over all scores, subject to the receiver best-replying.

\begin{remark}\label{prop:diagonal}
\textsl{Score $s_d$ is optimal for:\footnote{Score $s_d$ is the unique  optimal score up to an inconsequential relabeling of the messages.}
\begin{equation}\label{eq:cond1}
\frac{1}{f(0,0)}+\frac{1}{f(1,1)}<\frac{1}{f(1,0)}+\frac{1}{f(0,1)},
\end{equation}
and is  not an equilibrium score if:
\begin{equation}\label{eq:cond2}
f(1,0)<(\sqrt{2}-1)f(0,1).
\end{equation}
For $f(\theta)=\frac{1-\epsilon}{3}$ for $\theta\neq (1,0)$ and $f(1,0)= \epsilon $ for some $\epsilon\in(0,\frac{3}{\sqrt{2}}-2)$, both \eqref{eq:cond1} and \eqref{eq:cond2} hold.}
\end{remark}

\begin{proof}

\begin{figure}[h]
\begin{minipage}{.7\textwidth}
\centering
\begin{tikzpicture}[scale=1.5]
    \fill (0,0) circle (2pt);
    \fill (1,0) circle (2pt);
    \fill (0,1) circle (2pt);
    \fill (1,1) circle (2pt);
  \fill[fill= lightgray, opacity=0.5,rotate=45] (.7,0) ellipse (.3 and .9);
   \draw[  opacity=0.5] (0,0) ellipse (.25 and .25);
     \fill[fill=darkgray,opacity=0.5] (1,1) ellipse (.25 and .25);
\end{tikzpicture}\hspace{10pt}
\begin{tikzpicture}
    \draw (0,0) -- (0,2.8);
\end{tikzpicture}
\hspace{10pt}
\begin{tikzpicture}[scale=1.5]
    \fill (0,0) circle (2pt);
    \fill (1,0) circle (2pt);
    \fill (0,1) circle (2pt);
    \fill (1,1) circle (2pt);
  \fill[fill= lightgray, opacity=0.5,rotate=-45] (0,.7) ellipse (.3 and .9);
   \fill[fill=darkgray, opacity=0.5] (0,1) ellipse (.25 and .25);
     \draw[ opacity=0.5] (1,0) ellipse (.25 and .25);
\end{tikzpicture}

\begin{tikzpicture}
    \draw (0,0) -- (2.9,0);
    \draw (3.4,0) -- (6.4,0);
\end{tikzpicture}

\vspace{10pt}

\hspace{8pt}\begin{tikzpicture}[scale=1.5]
    \fill (0,0) circle (2pt);
    \fill (1,0) circle (2pt);
    \fill (0,1) circle (2pt);
    \fill (1,1) circle (2pt);
    
    \fill[lightgray, opacity=0.5] (0,0.5) ellipse (0.3 and .9);
    \draw[ opacity=0.5] (1,0.5) ellipse (0.3 and .9);
\end{tikzpicture}\hspace{10pt}
\begin{tikzpicture}
    \draw (0,0) -- (0,2.8);
\end{tikzpicture}
\hspace{10pt}
\begin{tikzpicture}[scale=1.5]
    \fill (0,0) circle (2pt);
    \fill (1,0) circle (2pt);
    \fill (0,1) circle (2pt);
    \fill (1,1) circle (2pt);
    \fill[lightgray, opacity=0.5] (0.5,0) ellipse (0.9 and .3);
    \draw[ opacity=0.5] (.5,1) ellipse (0.9 and .3);
\end{tikzpicture}
\end{minipage}
\begin{minipage}{.1\textwidth}
\centering \hspace{-100pt}
\begin{tikzpicture}
    \draw[thick] (-0.2, -1.2) rectangle (1.7, 0.4);
    \fill[gray]  (0,0) rectangle (0.2,0.2);
    \node[right] at (0.3,0.1) {$m=3$};
    \fill[lightgray] (0,-0.5) rectangle (0.2,-0.3);
    \node[right] at (0.3,-0.4) {$m=2$};
    \draw (0,-1) rectangle (0.2,-0.8);
    \node[right] at (0.3,-0.9) {$m=1$};
\end{tikzpicture}
\end{minipage}
\caption{Some scores in $\Theta=\{0,1\}^2$}\label{Scores}
\end{figure}

Let $s_{D}$, $s_{v}$ and $s_{h}$ denote, respectively, the scores  depicted in the top-right, bottom-left and bottom-right panel of \cref{Scores}. 
  It is easy  to verify that any optimal score must be identical, up to a relabeling of messages, to a score in the  set $\{s_d,s_D,s_h,s_v\}$. Let the  payoffs associated with $s_d$, $s_D$, $s_h$ and $s_v$  be respectively, $u_d$, $u_D$, $u_h$ and $u_v$ so that:
\begin{align*}
    u_{D}&:=-2 g(f(0,0),f(1,1)),\\
    u_{d}&:=-2 g(f(1,0),f(0,1)),\\ 
    u_{v}&:=- g(f(0,0),f(0,1)) -g(f(1,0),f(1,1)) ,\\
    u_{h}&:=- g(f(0,0),f(1,0)) -g(f(0,1),f(1,1)),
\end{align*}
where $g(x,y):=\frac{xy}{x+y}$.

Condition \eqref{eq:cond1} is equivalent to $g(f(1,0),f(0,1))<g(f(0,0),f(1,1))$, i.e., $u_{d}>u_{D}$.

Let $a:=f(0,0)$, $b:=f(1,1)$, $c:=f(1,0)$ and $d:=f(0,1)$. Condition \eqref{eq:cond1} can be written as:
\begin{equation}\label{eq:harm}
\frac{1}{a}+\frac{1}{b}<\frac{1}{c}+\frac{1}{d}.
\end{equation}
We show that \eqref{eq:harm} implies \begin{equation}\label{eq:harm2}\frac{ac}{a+c} + \frac{bd}{b+d}
  \;>\; \frac{2cd}{c+d}.
  \end{equation}
For any positive reals $p$ and $q$, the Arithmetic Mean-Harmonic Mean inequality gives
\[
  \frac{1}{p} + \frac{1}{q} \;\geq\; \frac{4}{p+q}.
\]
Set $p = \dfrac{1}{a}+\dfrac{1}{c}$ and $q = \dfrac{1}{b}+\dfrac{1}{d}$. Then
\begin{equation}\label{eq:step1}
  \frac{ac}{a+c} + \frac{bd}{b+d}
  \;=\; \frac{1}{p} + \frac{1}{q}
  \;\geq\; \frac{4}{p+q}
  \;=\; \frac{4}{\dfrac{1}{a}+\dfrac{1}{b}+\dfrac{1}{c}+\dfrac{1}{d}}.
\end{equation}
From \eqref{eq:harm},
\begin{equation}\label{eq:step2}
  \frac{4}{\dfrac{1}{a}+\dfrac{1}{b}+\dfrac{1}{c}+\dfrac{1}{d}}
  \;>\; \frac{4}{2\!\left(\dfrac{1}{c}+\dfrac{1}{d}\right)}
  \;=\; \frac{2}{\dfrac{1}{c}+\dfrac{1}{d}}
  \;=\; \frac{2cd}{c+d}.
\end{equation}
Conditions \eqref{eq:step1} and \eqref{eq:step2} together give us \eqref{eq:harm2}, which in turn is equivalent to $u_{d}>u_{h}$. 

The same argument, can be used to show \eqref{eq:harm} implies \begin{equation}\label{eq:harm3}\frac{ad}{a+d} + \frac{bc}{b+c}
  \;>\; \frac{2cd}{c+d},
  \end{equation}
which  is equivalent to $u_{d}>u_{v}$. We conclude that $s_{d}$ is the unique optimal score.

Consider now a PBE such that $\mu(\theta)=s_d(\theta)$ for all $\theta$. In such a PBE, $\mu(0,0)=1$, $\mu(0,1)=\mu(1,0)=2$ and $\mu(1,1)=3$; $\alpha(1)=(0,0)$,  $\alpha(2)=(\frac{f(1,0)}{f(1,0)+f(0,1)},\frac{f(0,1)}{f(1,0)+f(0,1)})$ and $\alpha(3)=(1,1)$.
Note that $u(\alpha(3),(1,0))=u(\alpha(1),(1,0))=-1$ hence 
\[
     u(\alpha(2),(1,0)) \geq u(\alpha(1),(1,0))\Leftrightarrow u(\alpha(2),(1,0)) \geq u(\alpha(3),(1,0))\Leftrightarrow 
     \frac{f(1,0)}{f(0,1)} \geq \sqrt{2}-1,
\]
 \Cref{eq:cond2} therefore implies that $s_d$ is not an equilibrium score.

The last part of the remark is immediate.
\end{proof}

\section{Proofs of \Cref{subsec:existencecommitment}}\label{app:proofProp:existence}

The results in this subsection hold for a large class of convex loss functions. We assume throughout that sender and receiver share the payoff function
\[
    u(a,\theta):=-c(a-\theta),
\]
where the loss function $c:\reals^{n}\to\reals$ is convex, $c(0)=0$ and $c(x)>0$ for every $x\neq 0$, i.e.   $a=\theta$ is the unique optimal action for both players. The following assumption ensures that expected losses are finite for every action:
\begin{assumption}\label{a:finite}
\[
   \E\bigl[c(2\tilde\theta)\bigr]<\infty \text{ and }\E\bigl[c(-2\tilde\theta)\bigr]<\infty,
\]
\end{assumption}

In order to prove \Cref{prop:existence}, we first record two ``technical'' lemmas. The first follows from convexity of $c$ and the
requirement that $c(x)>0$ for $x\neq0$; the second combines convexity with
\Cref{a:finite}.

\begin{lemma}\label{lem:coercive}
There exists $\kappa>0$ such that $c(x)\geq\kappa\|x\|$ for every
$x\in\reals^{n}$ with $\|x\|\geq1$. 
\end{lemma}

\begin{proof}
Let $\kappa:=\min_{x:\|x\|=1}c(x)$. As $c$ is convex and real-valued, it is
continuous; since $c(x)>0$ for $x\neq 0$ and the unit sphere is compact, $\kappa$
is well defined and strictly positive. Fix $x$ with $\|x\|\geq 1$ and let
$v:=x/\|x\|$. Then
\[
    \kappa\ \leq\ c(v)
    =c\Bigl(\tfrac{1}{\|x\|}\,\|x\|v+\bigl(1-\tfrac{1}{\|x\|}\bigr)\,0\Bigr)
    \ \leq\ \tfrac{1}{\|x\|}\,c(\|x\|v)=\tfrac{1}{\|x\|}\,c(x),
\]
where the first inequality holds because $\|v\|=1$, and the second holds because $c$ is convex, $c(0)=0$ and
$\tfrac{1}{\|x\|}\in(0,1]$. Hence $c(x)\geq\kappa\|x\|$.
\end{proof}

\begin{lemma}\label{lem:dominate}
For every $M>0$ there is a $C_M<\infty$ such that
\[
    \sup_{a:\|a\|\leq M}c(a-\theta)\ \leq\ C_M+\tfrac12\,c(-2\theta)
    \qquad\text{for every }\theta\in\reals^{n},
\]
and the function
$\theta\mapsto C_M+\tfrac12c(-2\theta)$ is integrable.
\end{lemma}

\begin{proof}
Let $C_M:=\tfrac12\max_{a:\|a\|\leq M}c(2a)$.  As $c$ is
continuous and $\{a:\|a\|\leq M\}$ is compact, then $C_M$ is well defined. For every $\theta$ and every $a$ with $\|a\|\leq M$,
\[
    c(a-\theta)=c\bigl(\tfrac12(2a)+\tfrac12(-2\theta)\bigr)
    \leq\tfrac12c(2a)+\tfrac12c(-2\theta)\leq C_M+\tfrac12c(-2\theta), 
\]
where the first inequality holds by convexity of $c$, and the second by construction of $C_M$.
As the bound  $C_M+\tfrac12c(-2\theta)$ is independent of
$a$, it also bounds the supremum of $c(a-\theta)$ over $a$ with $\|a\|\leq M$. 
  Finally,
\Cref{a:finite} implies that the function
$\theta\mapsto C_M+\tfrac12c(-2\theta)$ is integrable.
\end{proof}

The following sequence of lemmas establishes that there always exists an equilibrium  with two messages.  

Let $v(a^{I},a^{II})$ denote the expected loss when the receiver is restricted
to the two actions $a^{I}$ and $a^{II}$, and, for each state, the receiver is induced to take the action the players prefer:
\[
    v(a^{I},a^{II}):=\int_{\Theta}\min\bigl\{c(a^{I}-\theta),c(a^{II}-\theta)\bigr\}\,dF.
\]

\begin{lemma}\label{lem_vcontinuous} The function $v$ is continuous.
\end{lemma}

\begin{proof}
Take a convergent sequence $(a^{I,k},a^{II,k})\to(a^{I},a^{II})$. Being
convergent, the sequences $(a^{I,k})$ and $(a^{II,k})$ are bounded, say
$\|a^{I,k}\|\leq M$ and $\|a^{II,k}\|\leq M$ for every $k$, so by
\Cref{lem:dominate} there is a $C_M<\infty$ such that, for every $k$,
\[
    \min\bigl\{c(a^{I,k}-\theta),c(a^{II,k}-\theta)\bigr\}\leq C_M+\tfrac12c(-2\theta),
\]
where $C_M+\tfrac12c(-2\theta)$ is an integrable function of $\theta$ and is independent of $k$. Since $c$ and $(x,y)\mapsto\min\{x,y\}$
are continuous,
\[
    \min\{c(a^{I,k}-\theta),c(a^{II,k}-\theta)\}\to\min\{c(a^{I}-\theta),c(a^{II}-\theta)\}
    \quad\text{for each }\theta,
\]
and Lebesgue's dominated convergence theorem gives
$v(a^{I,k},a^{II,k})\to v(a^{I},a^{II})$. Hence $v$ is continuous.
\end{proof}

We establish next that the infimum of $v$ is attained by two \textsl{separate} actions.

\begin{lemma}\label{lem:maximization}
There exist $a^{I*},a^{II*}\in\reals^{n}$ such that $a^{I*}\neq a^{II*}$ and 
\begin{equation}\label{eq:2signals}
    v(a^{I*},a^{II*}) =\inf_{a^{I},a^{II}}v(a^{I},a^{II}).
\end{equation}
\end{lemma}

\begin{proof}
The function $v(a^{I},a^{II})$ is bounded below by $0$ and therefore an infimum
exists, say $\underline v$. Let 
$$V(a):=v(a,a)$$ 
denote
the expected loss when the receiver is restricted to the single action $a$, and let
$\bar v:=\inf_{a}V(a)$. As $v$ is continuous, so is $V$. Both $\underline v$ and
$\bar v$ are finite: $\max\{\underline v,\bar v\}\leq v(0,0)<\infty$, where the
second inequality follows from \Cref{lem:dominate}.

We first show that $\underline v<\bar v$. This requires showing that $\bar v$ is
attained.

Let the sequence $(a^{k})$ satisfy $V(a^{k})\to\bar v$ and suppose towards a contradiction that it is unbounded.
This sequence must have a subsequence such that $\|a^{k}\|\to\infty$. For every
$\theta\in\Theta$, along this subsequence $c(a^{k}-\theta)\to\infty$
(see \Cref{lem:coercive}), so
\[
     \lim_{k\to\infty}c(a^{k}-\theta)=\liminf_{k\to\infty}c(a^{k}-\theta)=\infty
    \qquad\text{for every }\theta\in\Theta.
\]
We therefore have 
\[
     \int_{\Theta}\liminf_{k\to\infty}c(a^{k}-\theta)\,dF  =\infty.
\]
Since $c$ is non-negative, Fatou's lemma applies and gives
\[
    \int_{\Theta}\liminf_{k\to\infty}c(a^{k}-\theta)\,dF
    \ \leq\ \liminf_{k\to\infty}\int_{\Theta}c(a^{k}-\theta)\,dF,
\]
so that $\int_\Theta c(a^k-\theta)\,dF \to \infty$, i.e. $V(a^k)\to\infty$. Yet we have established that $\bar v<\infty$. This
contradiction proves that every minimizing sequence is bounded. Therefore, the sequence $(a^k)$ has a convergent subsequence to some $a^*$ and $V(a^{*})=\bar v$, by continuity
of $V$.

Pick any $a'\neq a^{*}$. Since $c(a'-a')=0<c(a^{*}-a')$, continuity of $c$ implies
that the open set $G:=\{\theta:c(a'-\theta)<c(a^{*}-\theta)\}$ contains a
neighborhood of $a'$, hence $\Pr(\tilde\theta\in G)>0$. Therefore
\[
    \underline v\ \leq\ v(a^{*},a')
    =\int_{\Theta}\min\{c(a^{*}-\theta),c(a'-\theta)\}\,dF
    \ <\ \int_{\Theta}c(a^{*}-\theta)\,dF=\bar v .
\]

Next, let $(a^{I,k},a^{II,k})$ satisfy $v(a^{I,k},a^{II,k})\to\underline v$. We want
to show the sequence is bounded. Suppose it is not. Then at least one of the two
coordinates is unbounded, so there is a subsequence along which
$\Vert a^{i,k}\Vert\to\infty$ for some $i\in\{I,II\}$; up to a relabeling, take $i=I$.
Along that subsequence, the sequence $(a^{II,k})$ is either unbounded, in which
case we pass to a further subsequence along which $\Vert a^{II,k}\Vert\to\infty$ as
well, or bounded. We consider the two cases in turn. 

If $\Vert a^{i,k}\Vert\to\infty$ for both $i=I,II$, then
$\min\{c(a^{I,k}-\theta),c(a^{II,k}-\theta)\}\to\infty$ for every $\theta$ by
\Cref{lem:coercive}, so, as in the previous paragraph, Fatou's Lemma gives $v(a^{I,k},a^{II,k})\to\infty$, which
does not converge to $\underline v$ (recall that $\underline v<\infty$), a
contradiction.

If instead $(a^{II,k})$ is bounded, it admits a convergent subsequence, with limit
$a^{II}$ say. Along that subsequence $c(a^{I,k}-\theta)\to\infty$ by
\Cref{lem:coercive} and $c(a^{II,k}-\theta)\to c(a^{II}-\theta)$ by continuity of
$c$, so
\[
    \min\{c(a^{I,k}-\theta),c(a^{II,k}-\theta)\}\to c(a^{II}-\theta)
    \qquad \forall\theta\in\Theta.
\]
Moreover $(a^{II,k})$ is bounded by some $M$, so by \Cref{lem:dominate}
\[
    \min\{c(a^{I,k}-\theta),c(a^{II,k}-\theta)\}\leq c(a^{II,k}-\theta)
    \leq C_M+\tfrac12c(-2\theta),
\]
Lebesgue’s dominated convergence theorem then
gives
\[
    v(a^{I,k},a^{II,k})\longrightarrow\int_{\Theta}c(a^{II}-\theta)\,dF
    =V(a^{II})\ \geq\ \bar v,
\]
contradicting $v(a^{I,k},a^{II,k})\to\underline v<\bar v$.

The sequence $(a^{I,k},a^{II,k})$ is therefore bounded and admits a convergent
subsequence. By continuity of $v$, the limit of that subsequence, call it
$(a^{I*},a^{II*})$, attains $\underline v$. Finally, a pair with
$a^{I*}=a^{II*}$ would give $v(a^{I*},a^{I*})=V(a^{I*})\geq\bar v>\underline v$,
so the two actions must differ.
\end{proof}

\begin{lemma}\label{lem:2pbe}
There exists a Perfect Bayesian Equilibrium in which the sender chooses a strategy $\mu:\Theta\rightarrow \{1,2\}$.
\end{lemma}

\begin{proof}
Let $a^{I*},a^{II*}$ be as in \Cref{lem:maximization} and define
\[
    \mu(\theta):=
    \begin{cases}
        1 & \text{if } c(a^{I*}-\theta)\leq c(a^{II*}-\theta),\\
        2 & \text{otherwise},
    \end{cases}
    \qquad \alpha(1):=a^{I*},\quad\alpha(m):=a^{II*}, 
    \text{ for } m\neq 1
\]
and let $\Theta_{1}:=\mu^{-1}(1)$ and $\Theta_{2}:=\mu^{-1}(2)$. 

By construction, $\alpha(\mu(\theta))$ is, at every $\theta$, whichever of the two actions the players prefer, so the sender does not have a profitable deviation to the ``other'' on-path message. As off-path messages induce $a^{II*}$,  the sender does not have a profitable deviation to off-path messages either.

By the argument used for $G$ in the proof of \Cref{lem:maximization}, applied to
$a^{I*}$ and $a^{II*}$, both $\Theta_{1}$ and
$\Theta_{2}$ have positive probability. Beliefs are therefore given by Bayes' rule on path,
and any off-path message is assigned a belief associated with   on-path message $2$.

Finally, it is immediate that the receiver does not have a profitable deviation. Hence $(\mu,\alpha)$ is a Perfect Bayesian Equilibrium.
\end{proof}

\Cref{prop:existence} is a corollary of \Cref{lem:2pbe}.
\begin{proof}[Proof of \Cref{prop:existence}]
 By \Cref{lem:maximization}, $a^{I*}\neq a^{II*}$. By the argument in the proof of \Cref{lem:maximization}, both $\mu^{-1}(1)$ and $\mu^{-1}(2)$ have positive probability,
hence  $\mu$ is non-constant. Moreover, $\mu$ is measurable, since $\mu^{-1}(1)=\{\theta:c(a^{I*}-\theta)\leq
c(a^{II*}-\theta)\}$ is closed by continuity of $c$, and it satisfies the IVP:
for any $\theta,\theta'$ and any continuous path joining them,
$[\mu(\theta'),\mu(\theta)]\cap\mu(\Theta)\subseteq\{1,2\}$, and both values are
attained at the endpoints of the path. Hence $\mu$ is a score, and \Cref{prop:existence} then follows from \Cref{lem:2pbe}.
\end{proof}

In order to prove \Cref{proposition:calendar}, we first establish that $v^*=0$. The proof proceeds as follows. We construct a family of scores, indexed by some $\eta$ and some $K$. We then show that as $\eta\rightarrow 0$ and $K\rightarrow \infty$ ``in the appropriate way" the  payoff associated to the score in case the receiver best responds converges to $0$.

We will refer interchangeably to the CDF coordinates of the state and
the state itself. The following lemma ensures that this is without loss.

For $d=1,...,n$, let $F_d$ denote the marginal distribution function of
$\tilde\theta_d$.

\begin{lemma}\label{lem:quantile}
For each $d$, $F_d$ is a continuous, strictly increasing bijection from $\reals$
onto $(0,1)$; the random variable $\tilde z_d:=F_d(\tilde\theta_d)$ is uniform on
$[0,1]$.
\end{lemma}

The proof of the lemma is standard (see, for instance, Theorem 2.1.10 in \citet{casella2002statistical}) and therefore omitted.

The first step towards building the desired family of scores is to divide $\Theta$ into cells. For any $K\in\mathbb N$, such that $K\ge 2$ and $j=1,\dots,K$, let
\[
    I_j:=\begin{cases}
        \bigl[\tfrac{j-1}{K},\tfrac{j}{K}\bigr), & \text{ for }j<K,\\[4pt]
        \bigl[\tfrac{K-1}{K},1\bigr], & \text{ for }j=K,
    \end{cases}
\]
 and for
$i=(i_1,\dots,i_n)\in\{1,\dots,K\}^{n}$ let
\[
    Q_i:=\bigl\{\theta:\ F_d(\theta_d)\in I_{i_d}\ \text{ for } d=1,\dots,n\bigr\}.
\]
We refer to  $Q_i$ as a \emph{cell}. Let the sender strategy $\mu$ be such that, for every 
$i$, 
\[\theta\in Q_i\rightarrow \mu(\theta)=\lambda (i), \qquad \lambda(i):=\sum_{d=1}^{n}(i_d-1)K^{\,d-1}.\]
Figure \ref{fig:cells} shows, for $n = 2$ and $K = 3$, the cells $Q_i$ and the message generated by $\mu$.  The figure is drawn in the  coordinates $(z_1,z_2)$.

\begin{figure}[htbp]
\centering
\begin{tikzpicture}[scale=5.0]
  \draw[thick] (0,0) rectangle (1,1);
  \draw (1/3,0) -- (1/3,1); \draw (2/3,0) -- (2/3,1);
  \draw (0,1/3) -- (1,1/3); \draw (0,2/3) -- (1,2/3);
  \foreach \i/\x in {1/{0.1667},2/{0.5},3/{0.8333}}
    \foreach \j/\y in {1/{0.1667},2/{0.5},3/{0.8333}}
      {\pgfmathtruncatemacro{\lab}{(\i-1)+3*(\j-1)}
       \node[font=\small] at (\x,\y) {$\lab$};}
  \node[below] at (0.5,-0.05) {$z_1$};
  \node[left] at (-0.05,0.5) {$z_2$};
\end{tikzpicture}
\caption{The cells $Q_i$ and function $\mu$.}
\label{fig:cells}
\end{figure}

The strategy $\mu$ is not a score, as the message ``jumps" across cells in ways that violate the IVP. To construct the desired family of scores, we modify the strategy by adding appropriate ``transition windows'' between cells.  

For any $K\in\mathbb N$, such that $K\ge 2$, and $\eta\in(0,\frac{1}{2K})$, define $\psi_{K,\eta}:[0,1]\to[0,K-1]$ by
\[
    \psi_{K,\eta}(z):=\sum_{j=1}^{K-1}
    \min\Bigl\{1,\ \max\Bigl\{0,\ 1+\tfrac{1}{\eta}\bigl(z-\tfrac{j}{K}\bigr)\Bigr\}\Bigr\}.
\]
 
Let
\[
    g_{K,\eta}(\theta):=\sum_{d=1}^{n}K^{\,d-1}\,\psi_{K,\eta}\bigl(F_d(\theta_d)\bigr).
\]
 For $x\in\reals$, let $\lfloor x\rfloor:=\max\{k\in\mathbb Z:k\leq x\}$. For a function $g:\Theta\to\reals$, we write $\lfloor g\rfloor$ for the function mapping each $\theta$ to $\lfloor g(\theta)\rfloor$. Our desired family of scores is then given by:  
  \[
 s_{K,\eta}:=\bigl\lfloor g_{K,\eta}\bigr\rfloor.
\]
Let
\[
    N_\eta:=\Bigl\{\theta:\ F_d(\theta_d)\in\bigl(\tfrac{j}{K}-\eta,\tfrac{j}{K}\bigr)
    \ \text{ for some } d\leq n,\text{ and }\ j\leq K-1\Bigr\}
\]
be the set of states for which at least one coordinate $d$ is such that $\psi_{K,\eta}(F_d(\theta_d))$ is not locally constant. Note that
\begin{equation}\label{eq:identify}
    \theta\in Q_i\setminus N_\eta
    \quad\Longrightarrow\quad
    s_{K,\eta}(\theta)=g_{K,\eta}(\theta)=\mu(\theta).
\end{equation}
Figure \ref{fig:gs} shows, for $n = 2$, $K = 3$, and $\eta=0.15$, the function $g_{K,\eta}$ (panels (a) and (b)) and the induced score $s_{K,\eta}$ (panels (c) and (d)).  The panels are, once again,  drawn  in   coordinates $ (z_1,z_2)$. In panel (a),  the union of the shaded areas gives $N_\eta$.  Panel (b) follows $g_{K,\eta}$ up the dashed line. Panel (c) shows the mapping from $z$ to messages induced by $s_{K,\eta}$. To help the comparison  with panel (a), panel (c) also shows the shaded areas from panel (a).  Panel (d) is the same slice as (b), but for the score. 
 \begin{figure}[htbp]
\centering
\begin{subcaptionblock}{0.46\textwidth}
\centering
\begin{tikzpicture}[scale=5.0]
  \def\p{0.183333}   
  \def\q{0.516667}   
  \fill[black!10] (\p,0) rectangle (1/3,1);
  \fill[black!10] (\q,0) rectangle (2/3,1);
  \fill[black!10] (0,\p) rectangle (1,1/3);
  \fill[black!10] (0,\q) rectangle (1,2/3);
  \draw[thick] (0,0) rectangle (1,1);
  \draw (1/3,0) -- (1/3,1); \draw (2/3,0) -- (2/3,1);
  \draw (0,1/3) -- (1,1/3); \draw (0,2/3) -- (1,2/3);
  \foreach \i/\x in {1/{0.0917},2/{0.425},3/{0.8333}}
    \foreach \j/\y in {1/{0.0917},2/{0.425},3/{0.8333}}
      {\pgfmathtruncatemacro{\lab}{(\i-1)+3*(\j-1)}
       \node[font=\small] at (\x,\y) {$\lab$};}
  \draw[dashed,thick] (0.035,0) -- (0.035,1);
  \node[below] at (0.5,-0.05) {$z_1$};
  \node[left] at (-0.05,0.5) {$z_2$};
\end{tikzpicture}
\caption{$g_{K,\eta}$ on the unit square.}
\label{fig:gsquare}
\end{subcaptionblock}
\hfill
\begin{subcaptionblock}{0.46\textwidth}
\centering
\begin{tikzpicture}[x=3.4cm,y=0.33cm]
  \draw[->] (-0.04,0) -- (1.10,0) node[right] {$z_2$};
  \draw[->] (0,-0.8) -- (0,7.4);
  \foreach \y in {1,2,4,5} \draw[dotted,gray] (0,\y) -- (1.05,\y);
  \foreach \y in {3,6} \draw[dashed,gray] (0,\y) -- (1.05,\y);
  \foreach \y in {1,2,3,4,5,6} \draw (0.008,\y) -- (-0.008,\y);
  \node[left,font=\small] at (-0.010,3) {$3$};
  \node[left,font=\small] at (-0.010,6) {$6$};
  \node[below left,font=\small] at (0,0) {$0$};
  \draw[very thick]
    (0,0) -- (0.18333,0) -- (1/3,3) -- (0.51667,3) -- (2/3,6) -- (1.05,6);
  \foreach \x/\l in {{1/3}/{\tfrac13},{2/3}/{\tfrac23},1/{1}}
    {\draw (\x,0.15) -- (\x,-0.15); \node[below,font=\small] at (\x,-0.17) {$\l$};}
\end{tikzpicture}
\caption{$g_{K,\eta}$ along the dashed line.}
\label{fig:gslice}
\end{subcaptionblock}
 
\vspace{1.2\baselineskip}
 
\begin{subcaptionblock}{0.46\textwidth}
\centering
\begin{tikzpicture}[scale=5.0]
  \def\p{0.183333}
  \def\q{0.516667}
  \def\a{0.233333}
  \def\b{0.283333}
  \def\c{0.566667}
  \def\d{0.616667}
  \fill[black!10] (\p,0) rectangle (1/3,1);
  \fill[black!10] (\q,0) rectangle (2/3,1);
  \fill[black!10] (0,\p) rectangle (1,1/3);
  \fill[black!10] (0,\q) rectangle (1,2/3);
  \draw[thick] (0,0) rectangle (1,1);
  \foreach \x in {1/3,2/3}{
    \draw (\x,0) -- (\x,\p);
    \draw (\x,1/3) -- (\x,\q);
    \draw (\x,2/3) -- (\x,1);}
  \foreach \y in {\a,\b,1/3,\c,\d,2/3}{
    \draw (0,\y) -- (\p,\y);
    \draw (1/3,\y) -- (\q,\y);
    \draw (2/3,\y) -- (1,\y);}
  \foreach \x/\X in {\p/{1/3}, \q/{2/3}}{
    \draw (\x,\a) -- (\X,\p);
    \draw (\x,\b) -- (\X,\a);
    \draw (\x,1/3) -- (\X,\b);
    \draw (\x,\c) -- (\X,\q);
    \draw (\x,\d) -- (\X,\c);
    \draw (\x,2/3) -- (\X,\d);}
  \foreach \i/\x in {1/{0.0917},2/{0.425},3/{0.8333}}
    \foreach \j/\y in {1/{0.0917},2/{0.425},3/{0.8333}}
      {\pgfmathtruncatemacro{\lab}{(\i-1)+3*(\j-1)}
       \node[font=\small] at (\x,\y) {$\lab$};}
  \node[font=\tiny] at (0.11,0.2583) {$1$};
  \node[font=\tiny] at (0.11,0.3083) {$2$};
  \node[font=\tiny] at (0.11,0.5917) {$4$};
  \node[font=\tiny] at (0.11,0.6417) {$5$};
  \draw[dashed,thick] (0.035,0) -- (0.035,1);
  \node[below] at (0.5,-0.05) {$z_1$};
  \node[left] at (-0.05,0.5) {$z_2$};
\end{tikzpicture}
\caption{$s_{K,\eta}$ on the unit square.}
\label{fig:ssquare}
\end{subcaptionblock}
\hfill
\begin{subcaptionblock}{0.46\textwidth}
\centering
\begin{tikzpicture}[x=3.4cm,y=0.33cm]
  \draw[->] (-0.04,0) -- (1.10,0) node[right] {$z_2$};
  \draw[->] (0,-0.8) -- (0,7.4);
  \foreach \y in {1,2,4,5} \draw[dotted,gray] (0,\y) -- (1.05,\y);
  \foreach \y in {3,6} \draw[dashed,gray] (0,\y) -- (1.05,\y);
  \foreach \y in {1,2,3,4,5,6} \draw (0.008,\y) -- (-0.008,\y);
  \node[left,font=\small] at (-0.010,3) {$3$};
  \node[left,font=\small] at (-0.010,6) {$6$};
  \node[below left,font=\small] at (0,0) {$0$};
  \draw[very thick] (0,0) -- (0.23333,0);
  \draw[very thick] (0.23333,1) -- (0.28333,1);
  \draw[very thick] (0.28333,2) -- (1/3,2);
  \draw[very thick] (1/3,3) -- (0.56667,3);
  \draw[very thick] (0.56667,4) -- (0.61667,4);
  \draw[very thick] (0.61667,5) -- (2/3,5);
  \draw[very thick] (2/3,6) -- (1.05,6);
  \foreach \x/\yl/\yh in
    {0.23333/0/1, 0.28333/1/2, 0.33333/2/3, 0.56667/3/4, 0.61667/4/5, 0.66667/5/6}
    {\draw[gray,thin] (\x,\yl) -- (\x,\yh);}
  \foreach \x/\l in {{1/3}/{\tfrac13},{2/3}/{\tfrac23},1/{1}}
    {\draw (\x,0.15) -- (\x,-0.15); \node[below,font=\small] at (\x,-0.17) {$\l$};}
\end{tikzpicture}
\caption{$s_{K,\eta}$ along the dashed line.}
\label{fig:sslice}
\end{subcaptionblock}
 
\caption{Functions $g_{K,\eta}$ and $s_{K,\eta}$, for $n=2$, $K=3$ and $\eta=0.15$.}
\label{fig:gs}
\end{figure}

\begin{lemma} For any $K\in\mathbb N$, such that $K\ge 2$, and $\eta\in(0,\frac{1}{2K})$, the function $s_{K,\eta}$ is a score. 
\end{lemma}
\begin{proof}
To prove the lemma we first show that for any continuous function 
  $h:\Theta\to\reals$ such that $\lfloor h\rfloor$ is non-constant,   $\lfloor h\rfloor$ is a score.

 Function $\lfloor h\rfloor$ is measurable by continuity of $h$, and non-constant by hypothesis, so only the
IVP has to be checked. Let $\theta,\theta'$ satisfy $\lfloor h(\theta)\rfloor>\lfloor h(\theta')\rfloor$, let $m\in[\lfloor h(\theta')\rfloor,\lfloor h(\theta)\rfloor]\cap \lfloor h\rfloor(\Theta)$, and let $\gamma:[0,1]\to\Theta$ be continuous with $\gamma(0)=\theta'$ and $\gamma(1)=\theta$. As $m$ is an integer, then $h(\theta')\ge m\Rightarrow \lfloor h(\theta')\rfloor\ge m $, hence $ \lfloor h(\gamma(0))\rfloor=\lfloor h(\theta')\rfloor=m$  and the IVP is
satisfied with $t=0$. Otherwise $h(\theta')<m\le \lfloor h(\theta)\rfloor\le h(\theta)$, and since $h\circ\gamma$ is continuous and $m$ lies between its endpoint values
$h(\theta')$ and $h(\theta)$, the intermediate value theorem yields $t\in[0,1]$
with $h(\gamma(t))=m$; as $m$ is an integer, $\lfloor h(\gamma(t))\rfloor=m$. Hence $\lfloor h\rfloor$ is indeed a score.

Each $F_d$ is continuous (see \Cref{lem:quantile}) and $\psi_{K,\eta}$ is continuous,
so $g_{K,\eta}$ is continuous.
As $g_{K,\eta}$ is non-constant, the first part of the proof ensures that $s_{K,\eta}$ is a score.
\end{proof}

We then show that as $\eta\rightarrow 0$ and $K\rightarrow \infty$ ``in the appropriate way" the   equilibrium payoff associated to the score converges to $0$. To do this, we consider the expected loss when the sender uses the score $s_{K,\eta}$ and the
receiver responds with a simple, generically non-optimal, strategy. This expected loss is evidently an upper bound for the equilibrium loss associated to the score.
 
For any $K$, let $\mathcal L(\eta)$
denote the expected loss when the sender uses the score $s_{K,\eta}$ and the
receiver responds with
\[
    a_{K,\eta}(m):=
    \begin{cases}
        \E[\tilde\theta\mid\tilde\theta\in Q_i], & \text{ if }m=\lambda(i)\ \text{ for some } i,\\
        0, & \text{otherwise},
    \end{cases}
\]
hence
\[
\mathcal L(\eta):=\E\bigl[c\bigl(a_{K,\eta}(s_{K,\eta}(\tilde\theta))-\tilde\theta\bigr)\bigr].
\]
Let furthermore $A_K(\theta):=\E[\tilde\theta\mid
\tilde\theta\in Q_i]$ for every $i$ and every $\theta\in Q_i$.

As we let the transition windows shrink, the score does as well as the strategy $\mu$.

\begin{lemma}\label{lem:eta-limit}
For every $K$,
\[
    \limsup_{\eta\to0}\ \mathcal L(\eta)\ \le\ \E\bigl[c(A_K-\tilde\theta)\bigr].
\]
\end{lemma}

\begin{proof}
By \eqref{eq:identify}, $a_{K,\eta}(s_{K,\eta}(\theta)) = A_K(\theta)$ off $N_\eta$, so:
\begin{align}
    \mathcal L(\eta)
    \ &=\
    \E\bigl[c(A_K-\tilde\theta) \mathbbm 1[\tilde\theta\notin N_\eta]\bigr]
    \;+\;
    \E\bigl[c\bigl(a_{K,\eta}(s_{K,\eta}(\tilde\theta))-\tilde\theta\bigr)
    \mathbbm 1[\tilde\theta\in N_\eta]\bigr]\nonumber\\
    &\le\
    \E\bigl[c(A_K-\tilde\theta)\bigr]
    \;+\;
    \E\bigl[c\bigl(a_{K,\eta}(s_{K,\eta}(\tilde\theta))-\tilde\theta\bigr)
    \mathbbm 1[\tilde\theta\in N_\eta]\bigr],\label{eq:split}
\end{align}
where the inequality follows from $c$ being non-negative.
Convexity of $c$ gives $c(a-\theta)\le\tfrac12c(2a)+\tfrac12c(-2\theta)$ for all
$a,\theta$, hence
\begin{align*}
    \E\bigl[c\bigl(a_{K,\eta}(s_{K,\eta}(\theta))-\theta\bigr)
      \mathbbm 1[\theta\in N_\eta]\bigr]
    \ \leq\ &
    \tfrac12\E\bigl[c\bigl(2a_{K,\eta}(s_{K,\eta}(\theta))\bigr)
      \mathbbm 1[\theta\in N_\eta]\bigr]\\
    &+\tfrac12\E\bigl[c(-2\theta)\mathbbm 1[\theta\in N_\eta]\bigr].
\end{align*}
As $a_{K,\eta}$ takes finitely many values, all independent of $\eta$,
\begin{equation}\label{eq:above}
    \E\bigl[c\bigl(2a_{K,\eta}(s_{K,\eta}(\theta))\bigr)
      \mathbbm 1[\theta\in N_\eta]\bigr]
    \ \leq\
    \max_i c\bigl(2\E[\theta\mid\theta\in Q_i]\bigr)\,\Pr(\theta\in N_\eta).
\end{equation}
By \Cref{lem:quantile} each $F_d(\theta_d)$ is uniform on $[0,1]$, so a
single window has marginal probability $\eta$; as there are $K-1$ windows for
each of the $n$ coordinates, a union bound gives
\begin{equation}\label{eq:ramp-small}
    \Pr(\theta\in N_\eta)\ \leq\ n(K-1)\eta .
\end{equation}
Hence, $\Pr(\theta\in N_\eta)\to0$ as $\eta\to0$, and the
right-hand side of \eqref{eq:above} vanishes; so does
$\E[c(-2\theta)\mathbbm 1[\theta\in N_\eta]]$, since $c(-2\theta)$ is integrable
by \Cref{a:finite}. The second term on the right of \eqref{eq:split} therefore
vanishes, while the first does not depend on $\eta$; taking $\limsup$ in
\eqref{eq:split} gives the claim.
\end{proof}

We let the number of cells grow along powers of two, so that each level refines
the previous one.

\begin{lemma}\label{lem:K-limit}
$\E\bigl[c(A_{2^{r}}-\tilde\theta)\bigr]\rightarrow 0$ as $r\to\infty$.
\end{lemma}

\begin{proof}

Let $\mathcal F_K:=\sigma\bigl(Q_i:i\in\{1,\dots,K\}^{n}\bigr)$. Since $\{Q_i\}_{i\in\{1,\dots,K\}^{n}}$ is a finite partition of $\reals^{n}$,
so $\E[\theta\mid\mathcal F_K]$ equals $\E[\theta\mid\theta\in Q_i]$ on each
$Q_i$, that is, $A_K$. Moreover, $\mathcal F_{2^{r}}\subseteq\mathcal F_{2^{r+1}}$.

The family of intervals defining the partition $\{Q_i\}$ for each $K$, $\{I^r_j\}_{r\in\mathbb{N},j=1,...,2^r}$ where $$I_j^r=\begin{cases}
        \bigl[\tfrac{j-1}{2^r},\tfrac{j}{2^r}\bigr), & \text{ for }j<2^r,\\
        \bigl[\tfrac{2^r-1}{2^r},1\bigr], & \text{ for }j=2^r,
    \end{cases}$$  generate $\mathcal B([0,1])$, so their
products generate $\mathcal B([0,1]^{n})$ and
$\sigma\bigl(\bigcup_r\mathcal F_{2^{r}}\bigr)=\sigma(\tilde z_1,\dots,\tilde z_n)
=\sigma(\tilde\theta)$, the last equality by \Cref{lem:quantile}. As
$\tilde\theta$ is integrable, Levy's upwards theorem gives $\E[\tilde\theta\vert \mathcal{F}_{2^r}]\rightarrow \tilde\theta$ almost surely as $r\rightarrow\infty$ and therefore $A_{2^r}\rightarrow\tilde\theta$, almost surely. Hence
$c(A_{2^{r}}-\tilde\theta)\to0$ almost surely, since $c$ is continuous with
$c(0)=0$.

By convexity of $c$ and the
conditional version of Jensen's inequality,
\[
    c(A_{2^{r}}-\tilde\theta)\ \le\ \tfrac12c(2A_{2^{r}})+\tfrac12c(-2\tilde\theta)
    \ =\ \tfrac12c\bigl(\E[2\tilde\theta\mid\mathcal F_{2^{r}}]\bigr)
        +\tfrac12c(-2\tilde\theta)
    \ \le\ \tfrac12\E\bigl[c(2\tilde\theta)\mid\mathcal F_{2^{r}}\bigr]
        +\tfrac12c(-2\tilde\theta).
\]
Both terms on the right are integrable by \Cref{a:finite}. The first is a
conditional expectation of the fixed integrable random variable $c(2\tilde\theta)$,
so the family $\bigl\{\E[c(2\tilde\theta)\mid\mathcal F_{2^{r}}]\bigr\}_r$ is
uniformly integrable; the second does not depend on $r$. Hence
$\{c(A_{2^{r}}-\tilde\theta)\}_r$ is uniformly integrable, and Vitali's convergence
theorem gives the claim.
\end{proof}

We are now ready to prove that $v^*=0$.

\begin{lemma}\label{lem:v0} $v^*=0$.
\end{lemma}
\begin{proof}

Fix $\varepsilon>0$. By \Cref{lem:K-limit} there is $r$ with
$\E\bigl[c(A_{2^{r}}-\tilde\theta)\bigr]<\varepsilon/2$; set $K:=2^{r}$. By
\Cref{lem:eta-limit},
\[
    \limsup_{\eta\to0}\ \mathcal L(\eta)\ \leq\ \E\bigl[c(A_{K}-\tilde\theta)\bigr]
    \ <\ \varepsilon/2,
\]
so there is $\eta\in\bigl(0,\tfrac{1}{2K}\bigr)$ with $\mathcal L(\eta)<\varepsilon$.

For this $K$ and $\eta$, score $s_{K,\eta}$, together
with the receiver strategy $a_{K,\eta}$, yields the payoff
$-\mathcal L(\eta)>-\varepsilon$. The receiver's best reply to $s_{K,\eta}$ does at
least as well, so $v^{*}>-\varepsilon$. Since $\varepsilon$ was arbitrary, $v^* \ge 0$. Since every score yields a payoff of at most $0$, then  $v^* = 0$.
\end{proof}

We now move on to prove that the equilibrium payoff is bounded below 0.

\begin{lemma}\label{prop:bounded}
  Let $S^*$ denote the set of equilibrium scores.  Then $\sup_{s\in S^*} \E[u(\alpha(s(\theta)),\theta)]<0$.
\end{lemma}

\begin{proof}

Fix three states $\theta_1,\theta_2,\theta_3\in\Theta$ with $\|\theta_i-\theta_j\|=\sqrt3$ for $i\neq j$, and let $C:=\operatorname{conv}\{\theta_1,\theta_2,\theta_3\}$.

We first show that for \textsl{any} equilibrium score $s$  there is $\hat\theta\in C$ with $\|\alpha(s(\hat\theta))-\hat\theta\|\ge\frac12$. We prove this by contradiction. Pick an equilibrium score $s$ with receiver strategy $\alpha$, and let $A:=\alpha(s(\Theta))$. 
Suppose  $\|\alpha(s(\theta))-\theta\|<\tfrac12$ for all $\theta\in C$. Let, up to a relabeling, $s(\theta_1)\le s(\theta_2)\le s(\theta_3)$.

If $s(\theta_i)=s(\theta_j)$ for some $i\neq j$, then $\alpha(s(\theta_i))=\alpha(s(\theta_j))=:a$ and
\[
\sqrt3=\|\theta_i-\theta_j\|\le\|\theta_i-a\|+\|a-\theta_j\|<1,
\]
a contradiction. Hence $s(\theta_1)<s(\theta_2)<s(\theta_3)$.

Let $T:=[\theta_1,\theta_3]\subset C$ and let $\gamma(t):=(1-t)\theta_1+t\theta_3$ be a continuous path in $\Theta$ from $\theta_1$ to $\theta_3$. Since $s(\theta_3)>s(\theta_1)$ and $m:=s(\theta_2)\in[s(\theta_1),s(\theta_3)]\cap s(\Theta)$, the IVP implies the existence of a $t^*$ such that $\theta':=\gamma(t^*)\in T$ satisfies $s(\theta')=s(\theta_2)$, and hence $\alpha(s(\theta'))=\alpha(s(\theta_2))$.

Since $\|\theta_i-\theta_j\|=\sqrt3$ for all $i\neq j$, the points $\theta_1,\theta_2,\theta_3$ are the vertices of an equilateral triangle with side $\sqrt3$. The distance from $\theta_2$ to the segment $[\theta_1,\theta_3]$  is the length of the segment $[\theta_2,\tfrac12(\theta_1+\theta_3)]$. By the Pythagorean theorem, this distance equals
$$\sqrt{(\sqrt3)^2-\left(\tfrac{\sqrt3}{2}\right)^2}=\tfrac32.$$
Therefore
\[
\tfrac32\le\|\theta'-\theta_2\|
\le\|\theta'-\alpha(s(\theta'))\|+\|\alpha(s(\theta_2))-\theta_2\|<\tfrac12+\tfrac12=1,
\]
where the second inequality is the triangle inequality, and the last inequality holds as $\theta',\theta_2\in C$. This is a contradiction, so $\hat\theta\in C$ with $\|\alpha(s(\hat\theta))-\hat\theta\|\ge\frac12$ exists.

We next show that the existence of a $\hat\theta\in C$ with $\|\alpha(s(\hat\theta))-\hat\theta\|\ge\frac12$ implies that the expected payoff associated with score $s$ is bounded below some strictly negative threshold, which does not depend on the  score $s$.

For $\delta>0$ let $\underline c(\delta):=\min_{\|y\|=\delta}c(y)$, which is well defined and strictly positive because $c$ is continuous, $c(y)>0$ for $y\neq0$, and the set $\{y:\|y\|=\delta\}$ is compact. We claim
\begin{equation}\label{eq:lowerbound}
c(x)\ \ge\ \underline c(\delta)\qquad\text{for all } \|x\|\ge\delta .
\end{equation}
Indeed, for $\|x\|\ge\delta$, convexity and $c(0)=0$ give
$c\big(\delta\tfrac{x}{\|x\|}\big)\le \tfrac{\delta}{\|x\|}\,c(x)\le c(x)$, and  $c\big(\delta\tfrac{x}{\|x\|}\big)\geq \underline c(\delta)$ by construction. For every $a\in A$, we have:
\[
c(a-\hat\theta)\ \ge\ c\big(\alpha(s(\hat\theta))-\hat\theta\big)\ \ge\ \underline c(1/2),
\]
where the first inequality follows from (S-IC) at $\hat\theta$, and the second from \eqref{eq:lowerbound} and $\|\alpha(s(\hat\theta))-\hat\theta\|\ge\frac12$.

Let moreover $\rho>0$ be such that $\|x\|\le\rho \Rightarrow c(x)<\underline c(1/2)$. 
This implies $\|a-\hat\theta\|>\rho$ for every $a\in A$. Hence, for $\theta\in B_{\rho/2}(\hat\theta)$ and $a\in A$,
\[
\|a-\theta\|\ \ge\ \|a-\hat\theta\|-\|\hat\theta-\theta\|\ >\ \rho/2 .
\]
In particular this holds for $a=\alpha(s(\theta))\in A$, so by \eqref{eq:lowerbound}
\[
u(\alpha(s(\theta)),\theta)=-c(\alpha(s(\theta))-\theta)\ \le\ -\underline c(\rho/2) .
\]

 Let
\[
p:=\min_{\theta\in C}\Pr\big[\tilde\theta\in B_{\rho/2}(\theta)\big].
\]
The function $\theta\rightarrow \Pr[\tilde\theta\in B_{\rho/2}(\theta)]$ is continuous and strictly positive. As  $C$ is compact, $p$ is well defined and  strictly positive. Since $u\le0$ everywhere,
\[
\E\big[u(\alpha(s(\tilde\theta)),\tilde\theta)\big]
\ \le\ -\underline c(\rho/2)\,\Pr\big[\tilde\theta\in B_{\rho/2}(\hat\theta)\big]
\ \le\ -\underline c(\rho/2)\,p\ <\ 0 .
\]
The bound $-\underline c(\rho/2) p$ does not depend on $(s,\alpha)$, so
$\sup_{s\in S^*}\E[u(\alpha(s(\tilde\theta)),\tilde\theta)]\le-\underline c(\rho/2) p<0$.
\end{proof}

\section{Proof of Proposition \ref{prop:linear}}\label{app:proofProp:linear}

\begin{proof}
    
    A score $s$ is an equilibrium score if for some system of beliefs derived from $s$ and $F$, $\alpha(m)=\E[\tilde\theta\vert s(\tilde\theta)=m]$\footnote{By the disintegration theorem (see e.g., \citet{kallenberg_2021}, Theorem 3.4), there exists a system of beliefs on $M$, $\{F(\cdot\vert m)\}_{m\in M}\subseteq \Delta(\Theta)$ such that $F(\cdot\vert m)$ is a regular conditional probability distribution derived from the distribution of $m$ induced by $s$ and $F$ and this system of beliefs is almost everywhere uniquely defined.} and $s$ satisfies the IC constraints when the receiver's strategy is $\alpha$. Let $M=s(\Theta)$, $A=\alpha(s(\Theta))$ and $\Theta(a)=\{\theta:\alpha(s(\theta))=a\}$. Let $||\theta||_\Phi=\sqrt{\theta^\top\Phi \theta}$. Observe that $u(a,\theta)\geq u(a',\theta)\Leftrightarrow ||\theta-a||_\Phi\leq ||\theta-a'||_\Phi$.

    For any two points, $x,y\in\reals^n$, with a slight abuse of notation, let $[x,y]=\text{conv}\,\{x,y\}$, $(x,y)=[x,y]\setminus\{x,y\}$ and  $[x,y)=[x,y]\setminus\{y\}$. Finally, let $\ell(x,y)$ be the line through the points $x,y$. 

    The following lemma will be used throughout the proof.
    \begin{lemma}\label{lemma:linear}
       Let $a,a'\in \reals^n$. If $u(a,\theta)\geq u(a',\theta)$, then $u(a,\theta')> u(a',\theta')$ for all $\theta'\in [a,\theta)$.
    \end{lemma}
    \begin{proof}
Take $\theta'\in [a,\theta)$. First assume that $a'\notin \ell(a,\theta')$.  Note that 
    \begin{align}
        ||a-\theta||_\Phi\leq ||a'-\theta||_\Phi<||\theta'-\theta||_\Phi+||a'-\theta'||_\Phi, \label{lin:2}
        \end{align}
        where the last inequality holds by the triangle inequality and is strict because $\theta$, $\theta'$ and $a'$ are not collinear. Note also that 
        \begin{align*}
        ||\theta'-\theta||_\Phi+||a-\theta'||_\Phi&=||a-\theta||_\Phi
       <||\theta'-\theta||_\Phi+||a'-\theta'||_\Phi&\\
        \Rightarrow\; u(a,\theta')&>u(a',\theta'),&
    \end{align*}
    where the equality holds as $a$, $\theta$ and $\theta'$ are collinear, and the first inequality follows from \eqref{lin:2}.

    If instead $a'\in \ell(a,\theta')$, we must have $a'\notin (a,\theta]$, otherwise $u(a,\theta)<u(a',\theta)$. But then, either $a\in (a',\theta')$ or $\theta\in (\theta',a')$. In both cases,   $u(a,\theta')>u(a',\theta')$.
    \end{proof}

    \begin{lemma}\label{lemma_collinear actions}
        In any equilibrium where the sender strategy is a score, all actions are collinear: there exists $T\subseteq \reals$ and $b,v\in\reals^n$, $v\neq 0$ such that $A=\{b+tv:t\in T\}$. Moreover, the set $T$ is closed.
    \end{lemma}

    \begin{proof}
        Suppose it is not the case and there is $a^1,a^2,a^3\in A$ that are not collinear. Without loss, assume that$$
        s(a^1)<s(a^2)<s(a^3).
        $$The messages need to be distinct since it is a strict best-reply for $\theta=a^i$ to induce action $a^i$.

        Let $\Theta^\geq(a,\Tilde{a}):=\{\theta:u(a,\theta)\geq u(\Tilde{a},\theta)\}$, and let $\Theta^>(a,\Tilde{a})$ and $\Theta^=(a,\Tilde{a})$ be defined in the same way. The set $\Theta^\geq(a,\Tilde{a})$ is the closed half-space defined by $2\theta^\top\Phi(\Tilde{a}-a)\leq \Tilde{a}^\top\Phi\Tilde{a}-a^\top\Phi a$. We argue now that the hyperplanes $\Theta^=(a^2,a^1),\Theta^=(a^2,a^3)$ intersect. To see this note that their normal vectors are $\Phi (a^1-a^2)$ and $\Phi(a^3-a^2)$. Since $a^1,a^2,a^3$ are not collinear, for all $\lambda\neq 0$, $\lambda (a^1-a^2)+(a^3-a^2)\neq 0$. Using that $\Phi$ is invertible and therefore defines an injective map, we obtain that for all $\lambda\neq 0$, $\lambda\Phi(a^1-a^2)\neq \Phi (a^3-a^2)$.  

        Let $\hat{\Theta}=\left(\Theta^\geq(a^2,a^1)\cap\Theta^\geq(a^2,a^3)\right)^c$. The set $\hat\Theta$ is the set of states that strictly prefer $a^1$ or $a^3$ over $a^2$. It is the union of two open half-spaces and is connected. Moreover, the set $\Theta(a^2)\subseteq \hat{\Theta}^c$ and both $a^1,a^3\in \hat{\Theta}$. Because $\hat{\Theta}$ is connected, we can find a path from $a^1$ to $a^3$ entirely in $\hat{\Theta}$. By the IVP, there is $\theta\in\hat{\Theta}$ such that $s(\theta)=s(a^2)$, a contradiction.

        To show that $T$ is closed, suppose it is not. Then there is $x\in\reals\setminus T$ such that for all $\epsilon>0$, $B_\epsilon(x)\cap T\neq \emptyset$. Then, $\theta=x$ does not have a best-reply, contradicting that $(s,\alpha)$ is an equilibrium.
    \end{proof}

    \begin{lemma}
        Let $a$ be an isolated point in $A$. Then $\inte\Theta(a)\neq\emptyset$. 
    \end{lemma}

    \begin{proof}
    
        Since $a$ is an isolated point in $A$, there is $\epsilon>0$, such that  $||a-a'||>\epsilon$ for all $a'\neq a$. Hence, there is $\eta>0$, such that $||a-a'||_\Phi>\eta$ for all $a'\neq a$, by the equivalence of norms in $\reals^n$. Therefore, for all $\theta$ such that $||a-\theta||_\Phi<\eta/2$, $\theta\in \Theta(a)$. Since $a\in\Theta(a)$, $a\in \inte\Theta(a)$.   \end{proof}

        Let $\Theta^\geq(a):=\{\theta:u(a,\theta)\geq u(a',\theta),\text{ for all }a'\in A\}$, $\Theta^>(a):=\{\theta:u(a,\theta)>u(a',\theta),\text{ for all }a'\in A\setminus \{a\}\}$ and let $\Theta^=(a):=\{\theta\in\Theta(a): u(a,\theta)=u(a',\theta),\text{ for some }a'\in A\setminus\{a\}\}$.

    \begin{lemma}\label{lem:noint} Let $a$ be a limit point in $A$. Then $\inte\Theta(a)=\emptyset$.
\end{lemma}
\begin{proof} 

We proceed by contradiction. Suppose $\inte\Theta(a)\neq\emptyset$. We have $\Theta^>(a)\subseteq\Theta(a)\subseteq\Theta^=(a)\cup\Theta^>(a)$.

The set $\Theta^=(a)$ has an empty interior as by \cref{lemma:linear}, if $u(a,\theta)=u(a',\theta)$, then $u(a,\theta')<u(a',\theta')$ for all $\theta'\in[a',\theta)$.  Therefore, $\Theta^>(a)$ has a non-empty interior.
    
We also have that $\Theta^=(a)$ is a subset of the $\text{boundary of }\Theta^>(a)$ as by \cref{lemma:linear}, if $u(a,\theta)\geq u(a',\theta)$, $u(a,\theta)>u(a',\theta)$ for all $\theta'\in[a,\theta)$. Therefore, for any $\theta\in \Theta^=(a)$ and open ball $B_\epsilon(\theta)$, $B_\epsilon(\theta)$ intersects with both $\Theta^>(a)$ and $\Theta\setminus\Theta^>(a)$.

Next, we argue that $\Theta^>(a)$ is convex. Since \begin{align*}
        u(\theta,a)> u(\theta,\Tilde{a})        \Leftrightarrow-2\theta^\top(\Tilde{a}-a)>a^\top\Phi a-\Tilde{a}^\top\Phi\Tilde{a},
    \end{align*}
    these inequalities are preserved under convex combinations of $\theta$, so $\Theta^>(a)$ is convex.

If the interior of $\Theta^>(a)$ is not empty and $\Theta^>(a)$ is convex, then the boundary of $\Theta^>(a)$ has measure zero in $\reals^n$ \citep[see e.g.,][]{lang1986}. This also implies that $\Theta^=(a)$ has measure zero. Moreover, since $\E[\theta\vert s(\theta)=m]=a$ for all $m\in\{m'\in M:\alpha(m')=a\}$, we have $$
\E[\theta\vert \theta\in \Theta(a)]=a.
$$ 
    Therefore, $$\E[\theta\vert \theta\in\Theta(a)]=\E[\theta\vert \theta\in \Theta^>(a)]=\E[\theta\vert \theta\in\inte\Theta^>(a)]=a,$$ which implies $a\in\inte\Theta^>(a)$ by the full support assumption and convexity of $\Theta^>(a)$. But then, because $a$ is a limit point of $A$, it means that $\inte\Theta^>(a)$ intersects with $A$ at a point different than $a$, i.e., there is a point $a'\in A\cap \inte\Theta^>(a)$, $a'\neq a$ and associated message $m'$ with $\alpha(m')=a'$ such that $u(a,a')< u(\alpha(m'),a')=0$. Therefore, type $\theta=a'$ has a profitable deviation, a contradiction. Hence, $\inte \Theta(a)=\emptyset$.    \end{proof}

By \cref{lemma_collinear actions}, there is $v,b\in\reals^n$, $v\neq 0$ and $T\subseteq\reals$ such that $A=\{b+tv:t\in T\}$. Define $\beta=\Phi v$.

\begin{lemma}\label{lemma:lim_lin}
    For any limit point $a$ in $A$, $\Theta(a)=\{\theta:\beta^\top\theta=\beta^\top a\}$.
\end{lemma}
    \begin{proof}
        For any $\theta\in \{\theta':\beta^\top\theta'=\beta^\top a\}$, any $a'=a+\lambda v\in A$ for some $\lambda\neq 0$, we have $$
        u(a,\theta)-u(a',\theta)=\lambda^2v^\top\Phi v>0.
        $$
        Therefore, $\{\theta:\beta^\top\theta=\beta^\top a\}\subseteq \Theta(a)$. Suppose there is $\theta'\in\Theta(a)\setminus\{\theta:\beta^\top\theta=\beta^\top a\}$. Without loss assume that $u(a,\theta')>u(a',\theta')$ for all $a'\neq a$, since otherwise by \cref{lemma:linear}, we can take any other state $\theta''\in(a,\theta')$ where $u(a,\theta'')>u(a',\theta'')$ for all $a'\neq a$. But then by similar arguments as in \cref{lem:noint}, $\text{conv}\left(\{\theta:\beta^\top\theta=\beta^\top a\}\cup\{\theta'\}\right)\subseteq \Theta(a)$. The set $\text{conv}\left(\{\theta:\beta^\top\theta=\beta^\top a\}\cup\{\theta'\}\right)$ has a non-empty interior, contradicting $\inte\Theta(a)=\emptyset$.
    \end{proof}

\begin{lemma}\label{lemma:iso_lin}
    Let $a\in A$ be an isolated point of $A$. For any $\theta\in\inte\Theta(a)$, if $\beta^\top\theta'=\beta^\top\theta$ then $\theta'\in\inte\Theta(a)$.
\end{lemma}

\begin{proof}
    Take any $\theta\in\inte\Theta(a)$ and $\theta'$ such that $\beta^\top\theta=\beta^\top\theta'$. Since $\theta\in\inte\Theta(a)$, there exists $\epsilon>0$ such that for all $\hat\theta\in B_\epsilon(\theta)$, $||\hat\theta-a||_\Phi< ||\hat\theta-\tilde a||_\Phi$.\footnote{We can take strict inequalities since the inequalities hold strictly at $\theta$, otherwise it would not be an interior point and if $||\hat\theta-a||_\Phi= ||\hat\theta-\tilde a||_\Phi$, then they hold strictly for $\hat\theta_\lambda=\lambda\theta+(1-\lambda)\hat\theta$ with $\lambda\in (0,1)$.} Take $\hat\theta'$ such that $||\theta'-\hat\theta'||<\epsilon$. We want to show that $\hat\theta'\in \Theta(a)$. 
    
    Let $\hat\theta=\theta+\hat\theta'-\theta'$. We have $\hat\theta\in B_\epsilon(\theta)$ since $||\theta-\hat\theta||=||\theta'-\hat\theta'||<\epsilon$. Moreover, $\beta^\top\hat\theta=\beta^\top\hat\theta'$ since$$
    \beta^\top\hat\theta=\beta^\top\theta+\beta^\top\hat\theta'-\beta^\top\theta'=\beta^\top\hat\theta',
    $$using  $\beta^\top\theta=\beta^\top\theta'$.
    
    We have that $||\hat\theta-a||_\Phi< ||\hat\theta-\tilde a||_\Phi$ for all $\Tilde{a}\in A\setminus\{a\}$, which is equivalent to $2\hat\theta^\top\Phi(\Tilde{a}-a)< \Tilde{a}^\top\Phi\Tilde{a}-a^\top\Phi a$ for any $\Tilde{a}\in A$.
    
    By \cref{lemma_collinear actions}, $\Tilde{a}-a=\lambda v$ for some $\lambda\neq 0$. Moreover, $\beta=\Phi v$. Therefore, we also have, $2\hat\theta'^\top\Phi(\Tilde{a}-a)<\Tilde{a}^\top\Phi\Tilde{a}-a^\top\Phi a$ for any $\Tilde{a}\in A$ since $\beta^\top\hat\theta=\beta^\top\hat\theta'$.     
\end{proof}

Recall that by \cref{lemma_collinear actions}, there is $v,b\in\reals^n$, $v\neq 0$ and $T\subseteq\reals$ such that $A=\{b+tv:t\in T\}$ and that $\beta=\Phi v$. Let $A_\ell$ be the set of limit points in $A$ and $A_I$ the set of isolated points in $A$. Let $\Theta^\dagger=\left(\cup_{a\in A_\ell}\Theta(a)\right)\cup\left(\cup_{a\in A_I}\inte\Theta(a)\right)$. Let $X=\{\beta^\top\theta:\theta\in\Theta^\dagger\}$. For any $\theta\in\Theta^\dagger$ and $\theta'$ with $\beta^\top\theta'=\beta^\top\theta$, we have $$
\theta'\in\Theta^\dagger\text{ and }\alpha(s(\theta'))=\alpha(s(\theta)).
$$Therefore, there is a function $h:X\rightarrow T$ such that $$
\alpha(s(\theta))=b+h(\beta^\top\theta)v.
$$We now show that $h$ is increasing. Take any $x_1<x_2$ in $X$ and let $\theta^i\in\Theta^\dagger$ be such that $\beta^\top\theta^i=x_i$. Let $a^i=b+h(x_i)v$. From the IC constraints, we get\begin{align*}
    &||\theta^1-a^1||_\Phi\leq ||\theta^1-a^2||_\Phi,\\
    &||\theta^2-a^2||_\Phi\leq ||\theta^2-a^1||_\Phi.
\end{align*}
Expanding and combining these inequalities, we obtain$$
2\theta^{1\top}\Phi(a^2-a^1)\leq a^{2\top}\Phi a^2-a^{1\top}\Phi a^1\leq 2\theta^{2\top}\Phi(a^2-a^1).
$$Since $\beta=\Phi v$ and $a^2-a^1=(h(\beta^\top\theta^2)-h(\beta^\top\theta^1))v$, we get $$
(h(\beta^\top\theta^2)-h(\beta^\top\theta^1))x_1\leq (h(\beta^\top\theta^2)-h(\beta^\top\theta^1))x_2 \Leftrightarrow 0\leq (h(\beta^\top\theta^2)-h(\beta^\top\theta^1))(x_2-x_1),
$$and therefore $h(x_2)\geq h(x_1)$.

We now show that $\Theta\setminus\Theta^\dagger$ has Lebesgue measure zero. To see this, note that $A_I$ is countable and that following arguments similar to the proof of \cref{lem:noint}, the boundary of $\Theta(a)$ has Lebesgue measure zero when $a\in A_I$. Therefore $\Theta\setminus\Theta^\dagger=\cup_{a\in A_I}(\Theta(a)\setminus\inte\Theta(a))$ has Lebesgue measure zero and probability zero. Moreover, $X$ is dense in $\reals$, otherwise there is an open set $\tilde X\subset \reals\setminus X$ which implies $\{\theta: \beta^\top\theta= x:x\in\tilde X\}\subset \Theta\setminus \Theta^\dagger$ has positive measure. 

Therefore, we can extend $h:X\rightarrow \reals$ to $\hat{h}:\reals\rightarrow\reals$ with$$
\hat{h}(x)=\begin{cases}
    h(x)&x\in X,\\
    \sup_{y\in X, y<x}h(y)&x\notin X.
\end{cases}
$$
The supremum is finite since $h$ is bounded by $h(x')$ for some $x'\in X$, $x'>x$. Moreover since $T$ is closed, $h(x)\in T$ for all $x\in\reals$.

Define the score $\hat{s}(\theta)=\hat{h}(\beta^\top\theta)$. Since $\text{Pr}[\theta\in\Theta^\dagger]=1$, we get $$
\alpha(s(\theta))=b+\hat{s}(\theta)v, \text{ almost surely.}
$$
Moreover, we have $\alpha(s(\theta))=\E[\tilde\theta\vert s(\theta)]$.
For all $\theta\in\Theta^\dagger$, the original score defines the new score through $\hat{s}(\theta)=\frac{v^\top(\alpha(s(\theta))-b)}{v^\top v}$.

We want to show that for each $\theta\in\Theta^\dagger$, $\E[\tilde\theta\vert \hat{s}(\tilde\theta)=\hat{s}(\theta)]=\alpha(s(\theta))$, for some regular conditional expectation.  Fix $\theta\in\Theta^\dagger$ and let $\hat{m}=\hat{s}(\theta)$. 
\begin{align*}
    \E[\tilde\theta\vert \hat{s}(\tilde\theta)=\hat{m}]&=\E_m[\E[\tilde\theta\vert s(\tilde\theta)=m, \hat{s}(\tilde\theta)=\hat{m}]\vert \hat{s}(\tilde\theta)=\hat{m}]\\
    &=\E_m[\E[\tilde\theta\vert s(\tilde\theta)=m]\vert \alpha(m)=b+\hat{m}v]\\
    &=\E_m[\alpha(m)\vert \alpha(m)=b+\hat{m}v]\\
    &=b+\hat{m}v=\alpha(s(\theta)),
\end{align*}
where the first equality follows from the tower property of conditional expectations, the second from the fact that $s$ defines $\hat{s}$ almost surely on $\Theta$ and that the set of $m$ sent by $\{\tilde\theta:\hat{s}(\tilde\theta)=\hat{m}\}$ is determined by $\alpha(m)=b+\hat{m}v$ and the third equality from the receiver's best reply. 

Therefore, $\hat{\alpha}(\hat{m})=b+\hat{m}v$ is a version of the conditional expectation $\E[\tilde\theta\vert \hat{s}(\tilde\theta)=\hat{m}]$ and for this version, $\E[\tilde\theta\vert \hat{s}(\tilde\theta)=\hat{s}(\theta)]=\alpha(s(\theta))$, for each $\theta\in\Theta^\dagger$. This fact establishes that the equilibrium $(\hat{s},\hat{\alpha})$ is payoff equivalent to $(s,\alpha)$.

It remains to show $\hat{s}$ satisfies the sender's IC constraints and the IVP.

To see it satisfies the IVP, take $\theta',\theta$ and $m\in (\hat{s}(\theta),\hat{s}(\theta'))\cap \hat{s}(\Theta)$ and a continuous path $\gamma:[0,1]\rightarrow\Theta$ with $\gamma(0)=\theta,\gamma(1)=\theta'$. We have to find $t^*$ such that $\hat{s}(\gamma(t^*))=m$. Because $m\in \hat{s}(\Theta)$, there is $\theta''$ such that $$
\hat{h}(\beta^\top\theta)<m=\hat{h}(\beta^\top\theta'')<\hat{h}(\beta^\top\theta').
$$
Since $\hat{h}$ is increasing and $\beta^\top\theta\neq\beta^\top\theta''\neq\beta^\top\theta'$, we must have $\beta^\top\theta<\beta^\top\theta''<\beta^\top\theta'$. Since the function $\beta^\top\gamma(t)$ is continuous, by the intermediate value theorem, there is $t^*\in (0,1)$ with $\beta^\top\gamma(t^*)=\beta^\top\theta''$. Therefore, $\hat{s}(\gamma(t^*))=m$.

To check the sender IC constraints, first observe that all $\theta\in\Theta^\dagger$ induce the same actions under $(\hat{s},\hat{\alpha})$ and $(s,\alpha)$. For $\theta\notin \Theta^\dagger$, take $x^n\in X$ such that $x^n<\beta^\top \theta:=$ and $x^n\rightarrow\beta^\top \theta$. Let $$
\theta^n=\theta+\frac{x^n-x}{||\beta||^2}\beta.
$$
Then $\theta^n\in\Theta^\dagger$ as $\beta^\top\theta^n=x^n$ and $\theta^n\rightarrow \theta$. Moreover, $\hat{\alpha}(\hat{s}(\theta^n))=b+h(\beta^\top\theta^n)\rightarrow b+\hat{h}(\beta^\top\theta)$.

Sender IC at $\theta^n$ gives $u(\hat{\alpha}(\hat{s}(\theta^n)),\theta^n)\geq u(a',\theta^n)$ for all $a'\in A$. Taking limits, using the continuity of $u$, we obtain $$
u(\hat{\alpha}(\hat{s}(\theta)),\theta)\geq u(a',\theta),\text{ for all }a'\in A.
$$\end{proof}

\section{Proof of Proposition \ref{prop:normal}}\label{app:proofProp:normal}

    For any strategy $s(\theta)=\beta^\top \theta$, we have the unconditional distribution over messages $m$ induced by the score $s$, $m\sim N(0,\sigma_s^2)$ where $\sigma_s^2=\beta^\top\Sigma\beta$. We also have that $\text{Cov}(\theta,m)=\Sigma\beta$. 
    
    The payoff of the sender can be rewritten, up to a constant, as $$
    -a^\top\Phi a+2a^\top\Phi\theta.
    $$
    Therefore, the ex-ante payoff, given that the best-reply to $m$ is $\alpha(m)=\frac{\Sigma\beta}{\beta^\top\Sigma\beta}m$, is \begin{align*}
    &\E[-\alpha(m)^\top\Phi\alpha(m)+2\alpha(m)^\top\Phi\theta]\\
    &=\E[-\frac{\beta^\top\Sigma}{\beta^\top\Sigma\beta}m\Phi\frac{\Sigma\beta}{\beta^\top\Sigma\beta}m+2\frac{\beta^\top\Sigma}{\beta^\top\Sigma\beta}m\Phi\theta]\\
    &=\frac{\beta^\top\Sigma \Phi\Sigma\beta}{\beta^\top\Sigma\beta},
    \end{align*}
    where the last equality follows from $\E[m^2]=\beta^\top\Sigma\beta$ and $\E[\theta m]=\Sigma\beta$. The matrix $\Sigma \Phi\Sigma$ is positive semidefinite and symmetric. Therefore, $\frac{\beta^\top\Sigma \Phi\Sigma\beta}{\beta^\top\Sigma\beta}$ is a generalized Rayleigh quotient \citep[see e.g.,][Chapter 15]{parlett1998} and its stationary points, up to a rescaling of $\beta$, of $\frac{\beta^\top\Sigma \Phi\Sigma\beta}{\beta^\top\Sigma\beta}$ are the eigenvectors of $\Sigma^{-1}(\Sigma \Phi\Sigma)=\Phi\Sigma$, i.e., the points $\beta$ such that there is $\lambda\in\reals$ such that $ \Phi\Sigma\beta=\lambda \beta$. Moreover, as generalized Rayleigh quotients attain a maximum and a minimum, one of the stationary points   corresponds to a maximizer and another to a minimizer.

    The equilibrium problem can be expressed as follows. Given a belief that the sender uses a linear strategy $\beta$, the receiver chooses $\alpha(m)=\frac{\Sigma\beta}{\beta^\top\Sigma\beta}m$. In equilibrium, the sender chooses a signal $m$ for each realization of $\theta$:$$
    \max_{m}-\frac{\beta^\top\Sigma m \Phi\Sigma\beta m}{(\beta^\top\Sigma\beta)^2}+2\frac{\beta^\top\Sigma m \Phi\theta}{\beta^\top\Sigma\beta}.
    $$The objective function is quadratic in $m$ and therefore the maximizer must satisfy the first-order condition: $$
    m=\beta^\top\Sigma\beta\frac{\beta^\top\Sigma \Phi}{\beta^\top\Sigma \Phi\Sigma \beta}\theta.
    $$Therefore, any equilibrium strategy must satisfy $$
    \beta^\top=\beta^\top\Sigma\beta\frac{\beta^\top\Sigma \Phi}{\beta^\top\Sigma \Phi\Sigma \beta} \Leftrightarrow \beta=\frac{\beta^\top\Sigma\beta}{\beta^\top\Sigma \Phi\Sigma \beta}\Phi\Sigma\beta.
    $$
    Take any equilibrium strategy $\beta$. From the equilibrium condition, $\beta$ is an eigenvector of $\Phi\Sigma$ with eigenvalue $\frac{\beta^\top\Sigma \Phi \Sigma\beta}{\beta^\top\Sigma\beta}$.

    Conversely, take an eigenvector $\beta$ of $\Phi\Sigma$, with eigenvalue $\lambda$. Plugging in the equilibrium condition, we get \begin{equation}\label{eq:equilibrium}
\beta=\beta^\top\Sigma\beta\frac{\Phi\Sigma\beta}{\beta^\top\Sigma \Phi\Sigma \beta}\Leftrightarrow \beta=\frac{\beta^\top\Sigma\beta}{\lambda \beta^\top\Sigma\beta}\lambda \beta,
    \end{equation}
where the equivalence follows from  $\Phi\Sigma\beta=\lambda \beta$ and $\beta^\top\Sigma \Phi=\lambda\beta^\top$. Equation (\ref{eq:equilibrium}) is satisfied and therefore $\beta$ is an equilibrium strategy.

\section{Proof of Proposition \ref{prop:normal_n=2} and \ref{prop:normal_n}}

We first prove for general $n$ (\cref{prop:normal_n}) and then show the specific result for $n=2$ (\cref{prop:normal_n=2}). 

Recall that $\beta^+$ solves $\max_{\beta\neq 0}\frac{\beta^\top\Sigma\Phi\Sigma\beta}{\beta^\top\Sigma\beta}$, which is equivalent to solving $\max_{\beta\neq 0}\frac{\beta^\top\Sigma\beta}{\beta^\top\Phi^{-1}\beta}$. To see this observe that the maximizer of the second generalized Rayleigh quotient is the eigenvectors associated with the largest eigenvalue of $(\Phi^{-1})^{-1}\Sigma$, the same solution as the original maximization problem.

We have$$
\frac{\beta^\top\hat{\Sigma}\beta}{\beta^\top\Phi^{-1}\beta}-\frac{\beta^\top\Sigma\beta}{\beta^\top\Phi^{-1}\beta}=\frac{2(\hat{\sigma}_{ij}-\sigma_{ij})\beta_i\beta_j}{\beta^\top\Phi^{-1}\beta}.
$$Let $\beta^+$ be optimal for $\Sigma$ and $\hat{\beta}^{+}$ for $\hat{\Sigma}$ and denote the largest eigenvalue of $\Phi\Sigma$ and $\Phi\hat{\Sigma}$ by $\lambda_+,\hat{\lambda}_+$. Normalizing the eigenvectors by $\beta^{+\top}\Phi^{-1}\beta^+$ and $\hat{\beta}^{+\top}\Phi^{-1}\hat{\beta}^{+}$ we get\begin{align*}
    &\hat{\lambda}_+\geq \lambda_++2(\hat{\sigma}_{ij}-\sigma_{ij})\beta^+_i\beta^+_j,\\
    &\lambda_+\geq \hat{\lambda}_++2(\sigma_{ij}-\hat{\sigma}_{ij})\hat{\beta}^{+}_i\hat{\beta}^{+}_j.
\end{align*}
Rearranging we get, $$
2(\hat{\sigma}_{ij}-\sigma_{ij})\hat{\beta}^{+}_i\hat{\beta}^{+}_j\geq 2(\hat{\sigma}_{ij}-\sigma_{ij})\beta^+_i\beta^+_j.
$$The proof for the worst equilibrium follows the same line. 

Now suppose $n=2$. Let $\lambda_+,\lambda_-$ be the largest and smallest eigenvalues. Let $\beta$ be an eigenvector associated with the largest eigenvalue. Recall that tr$(\Phi\Sigma)=\lambda_++\lambda_-$. Therefore, $\lambda_+\geq  \frac{\text{tr}(\Phi\Sigma)}{2}=\frac{\phi_1\sigma_1^2+\phi_2\sigma_2^2+2\phi_{12}\sigma_{12}}{2}\geq \min\{\phi_1\sigma_1^2,\phi_2\sigma_2^2\}+\phi_{12}\sigma_{12}$.

Suppose first that $\phi_1\sigma_1^2\geq \phi_2\sigma_2^2$.

Since $\Phi\Sigma\beta=\lambda_+\beta$, we get from the second row,$$
(\phi_{12}\sigma_1^2+\phi_2\sigma_{12})\beta_1+(\phi_{12}\sigma_{12}+\phi_2\sigma_2^2)\beta_2=\lambda_+\beta_2.
$$Rearranging and multiplying by $\beta_1$ on both sides, we obtain$$
(\phi_{12}\sigma_1^2+\phi_2\sigma_{12})\beta_1^2=(\lambda_+-\phi_{12}\sigma_{12}-\phi_2\sigma_2^2)\beta_1\beta_2.
$$The term multiplying $\beta_1\beta_2$ is strictly positive so, using that $\phi_i>0$ for $i=1,2$, $$
\beta_1\beta_2\geq 0\Leftrightarrow \sigma_{12}\geq -\phi_{12}\frac{\sigma_1^2}{\phi_2}= -\phi_{12}\max\{\frac{\sigma_1^2}{\phi_2},\frac{\sigma_2^2}{\phi_1}\}.
$$
Suppose instead that $\phi_1\sigma_1^2< \phi_2\sigma_2^2$. Then the first row of $\Phi\Sigma\beta=\lambda_+\beta$ gives$$
(\phi_1\sigma_1^2+\phi_{12}\sigma_{12})\beta_1+(\phi_{1}\sigma_{12}+\phi_{12}\sigma_2^2)\beta_2=\lambda_+\beta_1.
$$
By a similar reasoning as in the first case, we get that $$
\beta_1\beta_2\geq 0\Leftrightarrow \sigma_{12}\geq -\phi_{12}\frac{\sigma_2^2}{\phi_1}=-\phi_{12}\max\{\frac{\sigma_1^2}{\phi_2},\frac{\sigma_2^2}{\phi_1}\}.
$$
The proof for the eigenvectors associated with the smallest eigenvalue follows similar lines.

\section{Proof of Proposition  \ref{prop:comparative}}

We first prove the general case then show the case for $n=2$.

As observed in the proof of \cref{prop:normal_n}, the solution to the ex-ante best and worst linear score is also the solution $\frac{\beta^\top\Sigma\beta}{\beta^\top\Phi^{-1}\beta}-\text{tr}(\Phi\Sigma)$. Both terms are linear in each entry of $\Sigma$. Therefore $\max_{\beta\neq 0}\frac{\beta^\top\Sigma\beta}{\beta^\top\Phi^{-1}\beta}-\text{tr}(\Phi\Sigma)$ viewed as a function of $\sigma_{ij}$ or $\sigma_i^2$ is convex and $\min_{\beta\neq 0}\frac{\beta^\top\Sigma\beta}{\beta^\top\Phi^{-1}\beta}-\text{tr}(\Phi\Sigma)$ is concave.

Let $\beta$ be an ex-ante best linear score. By convexity, the value $v^+(\Sigma)$ is differentiable in $\sigma_i^2$ almost everywhere. Wherever it is differentiable, the derivative of  $v^+(\Sigma)$ is, by the envelope theorem \citep{milgrom_segal2002},$$
\frac{\partial v^+(\Sigma)}{\partial \sigma_i^2}=\frac{\beta_i^2}{\beta^\top\Phi^{-1}\beta}-\phi_i.
$$
To establish that the derivative is negative, note that we can write $\beta_i^2=((\Phi^{1/2}e_i)^\top\Phi^{-1/2}\beta)^2$, $\Phi^{1/2}$ exists and is invertible because $\Phi$ is positive definite and symmetric. By the Cauchy-Schwarz inequality, we obtain $$
\beta_i^2=((\Phi^{1/2}e_i)^\top\Phi^{-1/2}\beta)^2\leq ((\Phi^{1/2}e_i)^\top(\Phi^{1/2}e_i))((\Phi^{-1/2}\beta)^\top\Phi^{-1/2}\beta)=\phi_i\beta^\top\Phi^{-1}\beta.
$$ The same reasoning applies for the ex-ante worst linear score.

It remains to show the minimum and the maximum for the case of $n=2$. For that we claim that the bounds on $v^+$ and $v^-$ are obtained at $\beta=(1,0)^\top$ or $\beta'=(0,1)^\top$. Direct calculations at $\beta,\beta'$ gives \begin{align*}
    &v^+(\Sigma)\geq -(\phi_1\phi_2-\phi^2_{12})\min\left\{\frac{\sigma_1^2}{\phi_2},\frac{\sigma_2^2}{\phi_1}\right\},\\
    &v^-(\Sigma)\geq-(\phi_1\phi_2-\phi^2_{12})\max\left\{\frac{\sigma_1^2}{\phi_2},\frac{\sigma_2^2}{\phi_1}\right\}.
\end{align*}

To see these bounds are attained recall that, since tr$(\Phi\Sigma)=\lambda_++\lambda_-$, we have $v^+=-\lambda_-$ and $v^-=-\lambda_+$. 

Suppose first that $\phi_1\sigma_1^2\geq \phi_2\sigma_2^2$. Then, at $\sigma_{12}=-\phi_{12}\frac{\sigma_2^2}{\phi_1}=-\phi_{12}\min\{\frac{\sigma_1^2}{\phi_2},\frac{\sigma_2^2}{\phi_1}\}$, $\Phi\Sigma$ is triangular:$$
\Phi\Sigma=
\begin{pmatrix}
\phi_1\sigma_1^2+\phi_{12}\sigma_{12}
&
\phi_1\sigma_{12}+\phi_{12}\sigma_2^2\\
\phi_{12}\sigma_1^2+\phi_2\sigma_{12}
&
\phi_{12}\sigma_{12}+\phi_2\sigma_2^2
\end{pmatrix},
$$and its smallest eigenvalue is its smallest diagonal element: $(\phi_1\phi_2-\phi_{12}^2)\frac{\sigma_2^2}{\phi_1}$.

At $\sigma_{12}=-\phi_{12}\frac{\sigma_1^2}{\phi_2}=-\phi_{12}\max\{\frac{\sigma_1^2}{\phi_2},\frac{\sigma_2^2}{\phi_1}\}$, $\Phi\Sigma$ is also triangular and its largest eigenvalue is its largest diagonal element: $(\phi_1\phi_2-\phi_{12}^2)\frac{\sigma_1^2}{\phi_2}$.

A similar reasoning establishes that the bounds are attained when $\phi_1\sigma_1^2< \phi_2\sigma_2^2$.

\section{Value of Ignorance}\label{app:ignorance}

We show here, by way of an example, that sender and receiver \textsl{can} be better off if the sender is less informed.

Let $\Phi$ correspond to the $2\times 2$ identity matrix,  $\Theta=\{0,1,2\}\times\{0,2\}$,  $f(0,0)=f(1,2)=f(2,0)=\frac{\epsilon}{3}$ and $f(0,2)=f(1,0)=f(2,2)=\frac{1-\epsilon}{3}$, for some  $\epsilon<\frac{1}{6}$. As in \cref{sec:2X2}, we use the ``graph-theoretic'' definition of IVP. To apply it, we view $\{0,1,2\}\times\{0,2\}$ as a simple graph where each $\theta\in\Theta$ is a vertex and the edges link adjacent points, i.e., an edge links two vertices $\theta,\theta'$ such that either $\vert\theta_1-\theta_1'\vert=1$ and $\vert\theta_2-\theta_2'\vert=0$ or $\vert\theta_1-\theta_1'\vert=0$ and $\vert\theta_2-\theta_2'\vert=2$.

We consider the standard setting and a setting in which the sender does not observe   $\theta_2$ and therefore can only select strategies that assign the same message to any two states with the same $\theta_1$. In   \cref{fig:ignorance},   empty circles denote  low-probability states and filled circles high-probability ones. The figure  illustrates a score  that assigns   to each state $(\theta_1,\theta_2)$ a message equal to $\theta_1$. We refer to this as  \textsl{score} $s_{1}$.

\begin{figure}[h!] 
 \begin{minipage}{.8\textwidth}
 \centering
\begin{tikzpicture}[scale=1.5]
    \draw (0,0) circle (2pt);
    \fill (1,0) circle (2pt);
    \draw (2,0) circle (2pt);
    \fill (0,2) circle (2pt);
    \draw (1,2) circle (2pt);
    \fill (2,2) circle (2pt);
  \draw[ opacity=0.5] (0,1) ellipse (0.2 and 1.2);
    \fill[fill=lightgray, opacity=0.5] (1,1) ellipse (0.2 and 1.2);
      \fill[fill=darkgray, opacity=0.5] (2,1) ellipse (0.2 and 1.2);

    \node at (0,1) {0};
    \node at (1,1) {1};
    \node at (2,1) {2};

  \node[mark size=4pt, thick, color=black] at (0,1.8) {\pgfuseplotmark{x}};
   \node[mark size=4pt, thick, color=black] at (1,0.2) {\pgfuseplotmark{x}};
    \node[mark size=4pt, thick, color=black] at (2,1.8) {\pgfuseplotmark{x}};
\end{tikzpicture}
\end{minipage}\hspace{-2cm}
    \begin{minipage}{.1\textwidth}
 
\begin{tikzpicture}
    \draw[thick] (-0.2, -1.95) rectangle (2.1, 0.4);
    \fill[gray]  (0,0) rectangle (0.2,0.2);
    \node[right] at (0.3,0.1) {$m=2$};
    \fill[lightgray] (0,-0.5) rectangle (0.2,-0.3);
    \node[right] at (0.3,-0.4) {$m=1$};
    \draw (0,-1) rectangle (0.2,-0.8);
    \node[right] at (0.3,-0.9) {$m=0$};
    \node[mark size=4pt, thick, color=black] at (0.1,-1.6) {\pgfuseplotmark{x}};
    \node[right] at (0.3,-1.6) {$\E[\theta\vert m]$};
\end{tikzpicture}
\end{minipage}
\caption{\label{fig:ignorance}}\ 
\end{figure}

If the sender does not observe    $\theta_2$, then  score $s_{1}$ is an equilibrium strategy.\footnote{The argument relies on the observation that, since the sender observes only $\theta_1$, her posterior mean coincides with the receiver's action after the message $m=\theta_1$. Any other message therefore induces a strictly higher expected loss.}
We show that in the standard setting every equilibrium score  induces a larger loss than $s_{1}$.

The expected loss associated with $s_{1}$ is $4\epsilon(1-\epsilon)$. We show that every equilibrium score $s$ such that $s(\theta')=s(\theta'')$ for two high-probability states 
induces a loss larger than $4\epsilon(1-\epsilon)$.

Fix $\theta'=(0,2)$ and $\theta''=(2,2)$.
In any equilibrium in which $\mu(\theta)=s(\theta)$ for all $\theta$, the loss conditional on $\theta\in\{(0,2),(2,2)\}$ is minimized if $\alpha(s(\theta'))=(1,2)$. For such action, the loss, conditional on $\theta\in\{\theta',\theta''\}$, is  $1$. The expected loss associated with the score is thus at least 
\[\Pr(\theta\in\{\theta',\theta''\})\times 1=(1-\epsilon)\frac{2}{3}.\]
 and 
\[(1-\epsilon)\frac{2}{3}>4\epsilon(1-\epsilon)\Leftrightarrow \epsilon<\frac{1}{6}.\]

Similar calculations show that for $\theta=(1,0)$ and $\theta''=(2,2)$ or $\theta''=(0,2)$, the expected loss associated with the score is at least 
$(1-\epsilon)\frac{5}{6}$. As this loss is larger than  $(1-\epsilon)\frac{2}{3}$, the condition $\epsilon<\frac{1}{6}$ is sufficient  to ensure that every equilibrium score $s$ such that $s(\theta')=s(\theta'')$ for two high-probability states 
induces a loss larger than the one associated with $s_1$.

 We show next that every equilibrium score $s$ is such that $s(\theta')=s(\theta'')$ for two high-probability states.

We proceed by contradiction. Let score $s$ satisfy $s(0,2)\neq s(1,0)\neq  s(2,2) $. Without loss, let  $s(1,2)=1$.  We consider two cases: $s(1,0)=1$ and $s(1,0)\neq 1$.

If $s(1,0)=1$, let, without loss, $s(0,2)=0$ and  $s(2,2)=2$. In this case, the IVP requires $s(0,0)\neq 2$ and $s(2,0)\neq 0$. This in turn implies that - in any equilibrium such that $\mu(\theta)=s(\theta)$ for all $\theta$ - the receiver chooses  $\alpha(0)=(0,x)$, where $x\geq 2(1-\epsilon)$ and  $\alpha(1)=(y,z)$, where $z\leq 2\epsilon$. For any value of $x$, $y$ and $z$, in state $(1,2)$ the sender has a profitable deviation to report $m=0$ instead of $m=1$. So the score is not an equilibrium score.

If instead $s(1,0)\neq 1$, let, without loss, $s(1,0)=2$, $s(0,2)=0$ and $s(2,2)=1$.    The IVP requires $s(0,0)=s(1,2)=1$ and $s(2,0)\in\{1,2\}$. Regardless of whether $s(2,0)=1$ or $s(2,0)=2$, in any equilibrium such that $\mu(\theta)=s(\theta)$ for all $\theta$,   in state $(0,0)$  - the sender has a profitable deviation to report $m=0$ instead of $m=1$. 

We conclude that any equilibrium score in the standard setting is associated with a larger loss   than $s_{1}$: in this example, sender and receiver can be better off if the sender does not observe  the realization of $\theta_2$.
\end{document}